%% file: syritsyn-lat13.tex
\documentclass{PoS}

\usepackage{amsmath}
\usepackage{graphicx}

\title{Review of Hadron Structure Calculations on a Lattice}

\ShortTitle{Hadron Structure Calculations on a Lattice}

\author{\speaker{Sergey Syritsyn}\\%\thanks{A footnote may follow.}\\
        RIKEN-BNL Research Center, Brookhaven National Laboratory, Upton, NY, 11973, USA\\
        E-mail: \email{syritsyn@alum.mit.edu}}

\abstract{
  I present a review of the current status and the most recent achievements 
  in lattice QCD calculations of hadron structure.
  First, I overview the status and systematic uncertainties of nucleon structure 
  ``benchmark'' quantities that are well known from experiments and serve as a reference point 
  for the validity of lattice QCD methods.
  Next, I discuss the current status of calculations of form factors of the nucleon
  and highlight some recent results for other hadrons that are important for 
  understanding their internal dynamics.
  Wave functions of hadrons and their excitations may also be studied in lattice QCD,
  and I illustrate it with two recent examples of such calculations.
  Finally, I discuss in detail the state of calculations pertaining to the nucleon 
  spin puzzle.
}

\FullConference{31st International Symposium on Lattice Field Theory - LATTICE 2013\\
		July 29 - August 3, 2013\\
		Mainz, Germany}

\input{defs}

\begin{document}

\section{Benchmark quantities\label{sec:benchmark}}

With recent advances, the field of Lattice QCD is mature enough to provide reliable
information about the world of nuclear physics. 
The first major breakthrough was a successful calculation of the hadron 
spectrum~\cite{Durr:2008zz}. 
The next milestone that has nearly been reached is to verify
that lattice QCD captures internal dynamics and structure of hadrons correctly.
On this path, reproducing basic features about the most studied hadrons, the proton and
the neutron, is an essential milestone.
The nucleon structure observables discussed in this section have 
the least systematic ambiguity and stochastic error, and thus may be categorized 
as ``lattice QCD benchmark'' quantities.

\subsection{Nucleon axial charge}
\begin{figure}[ht!]
  \centering
  \begin{minipage}{.48\textwidth}
    \centering
    \includegraphics[width=\textwidth]{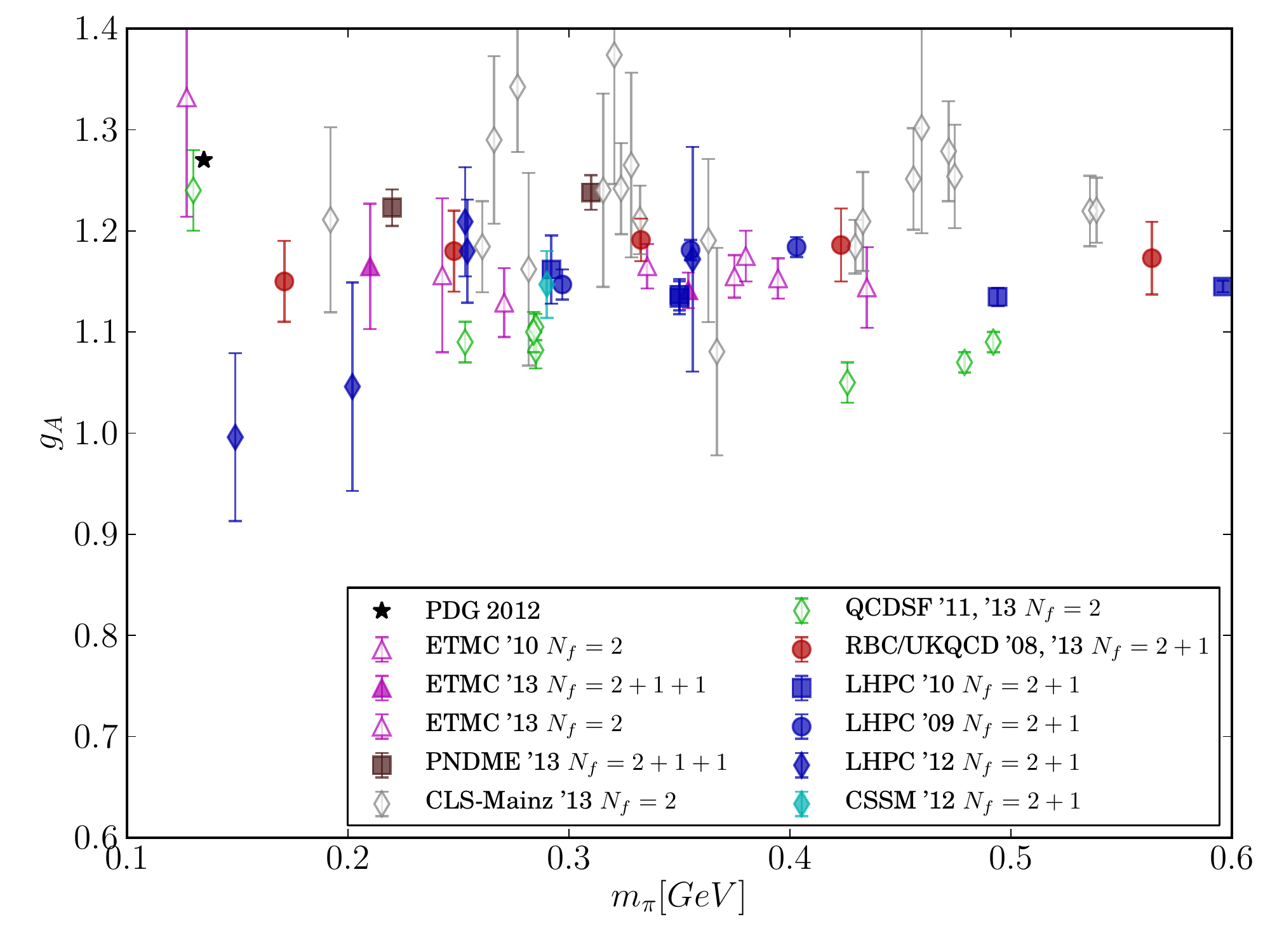}
    \caption{\label{fig:gA-summary}
      Summary of nucleon axial charge $g_A$ lattice results~\cite{Owen:2012ts,Alexandrou:2010hf,Alexandrou:2011aa,Alexandrou:2013joa,Bhattacharya:2013ehc,Capitani:2012gj,QCDSF:2011aa,Horsley:2013ayv,Ohta:2013qda,Bratt:2010jn,Green:2012ud}.}
  \end{minipage}~
  \hspace{.03\textwidth}~
  \begin{minipage}{.48\textwidth}
    \includegraphics[width=\textwidth]{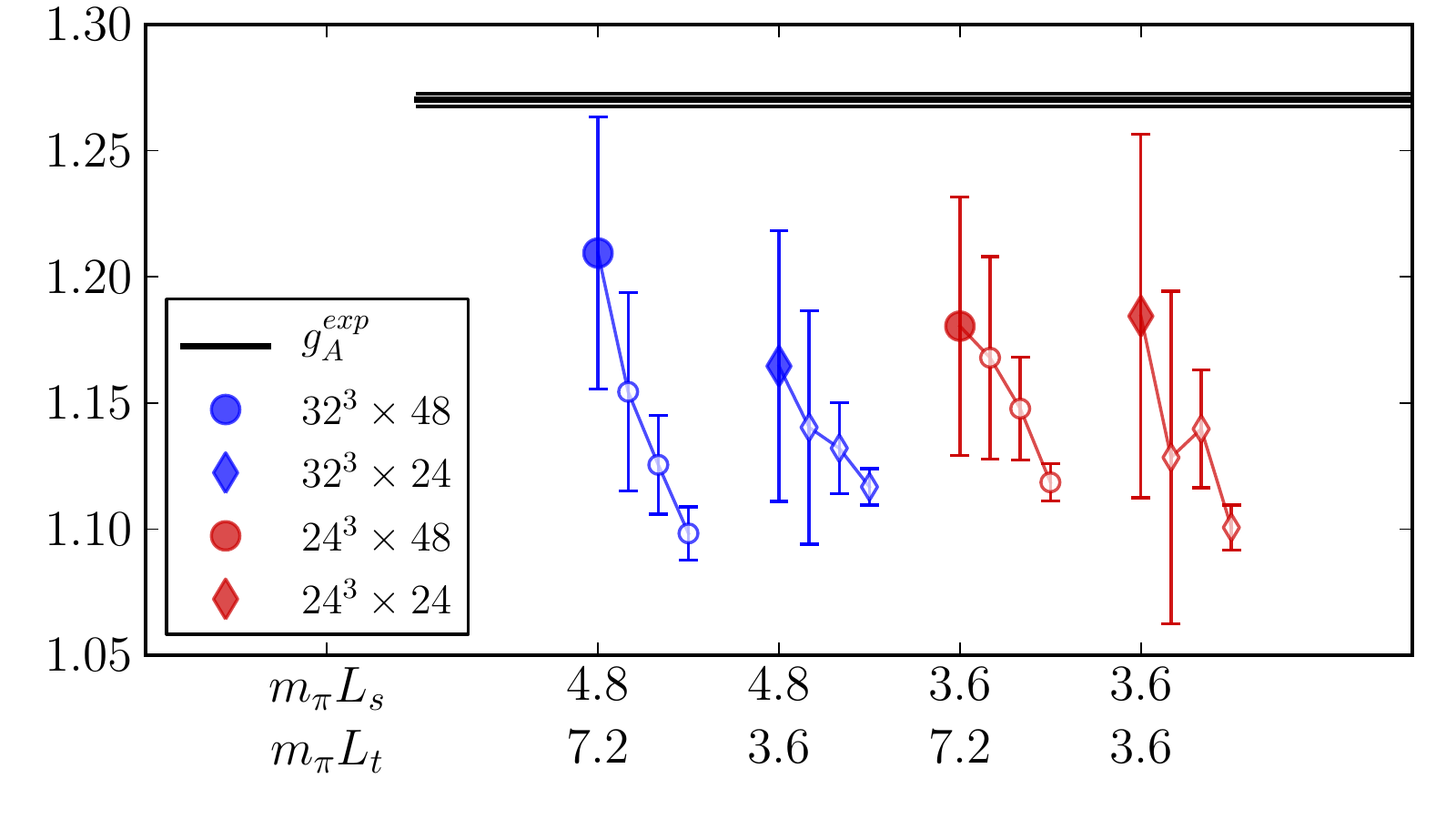}
    \caption{\label{fig:gA-vs-Ls-Lt}
      Detailed study of $g_A$ dependence on volume and temperature with Wilson fermions, 
      $m_\pi\approx250\text{ MeV}$~\cite{Green:2013hja}.}
  \end{minipage}
\end{figure}

The axial charge of the nucleon is an important quantity for the entire field of nuclear physics
as it governs the rate of $\beta$-decay and, for instance, the neutron half-life,
as a forward nucleon matrix element of the isovector axial-vector quark current,
\begin{equation}
\label{eqn:ga}
\langle N(P)|\bar{q}\gamma^\mu\gamma^5 q|N(P)\rangle 
  = g_A\, \bar{u}_P\gamma^\mu\gamma^5 u_P\,,
\end{equation}
where $u_p$ is the nucleon spinor, and which can be calculated without disconnected quark 
contractions ambiguity.
Historically, most attempts to calculate this quantity resulted in values 10-15\% 
below the experimental number, almost irrespective of the pion mass used.
Until recently, it was easy to ascribe this discrepancy to heavy pion masses used.
However, recent calculations with $m_\pi$ approaching the physical point apparently 
continue this trend.
Over the years, the deficiency has been ascribed to finite-volume effects
(FVE)~\cite{Yamazaki:2008py}, excited state
contributions~\cite{Owen:2012ts,Capitani:2012gj,Jager:2013kha} 
and finite-temperature effects~\cite{Green:2012ud}.
There is no convincing evidence that any of these sources is solely responsible for the
discrepancy; it is plausible that interplay of various systematic effects takes place, 
and with every collaboration using slightly different methods, meaningful comparisons are
complicated.
For instance, a better agreement was claimed upon removal of excited state contributions
with a variational~\cite{Owen:2012ts} and summation~\cite{Capitani:2012gj,Jager:2013kha}
methods, while others~\cite{Dinter:2011sg,Bhattacharya:2013ehc} did not detect 
significant excited state effects.
The pattern of finite volume dependence, initially claimed in~\cite{Yamazaki:2008py}, 
was not confirmed when other collaboration results were examined~\cite{Alexandrou:2013joa}.
Finite temperature dependence~\cite{Green:2012ud} was not confirmed in a subsequent detailed
study~\cite{Green:2013hja} at $m_\pi\approx250\text{ MeV}$ (Fig.~\ref{fig:gA-vs-Ls-Lt}).
Finally, there is a very encouraging agreement with experiment from the most recent 
study directly at the physical pion mass~\cite{Alexandrou:2013jsa}.
Hopefully, this preliminary result will be confirmed by other collaborations 
with careful evaluations of all systematic effects.

\subsection{Quark momentum fraction}

\begin{figure}[ht!]
  \begin{minipage}{.48\textwidth}
    \includegraphics[width=\textwidth]{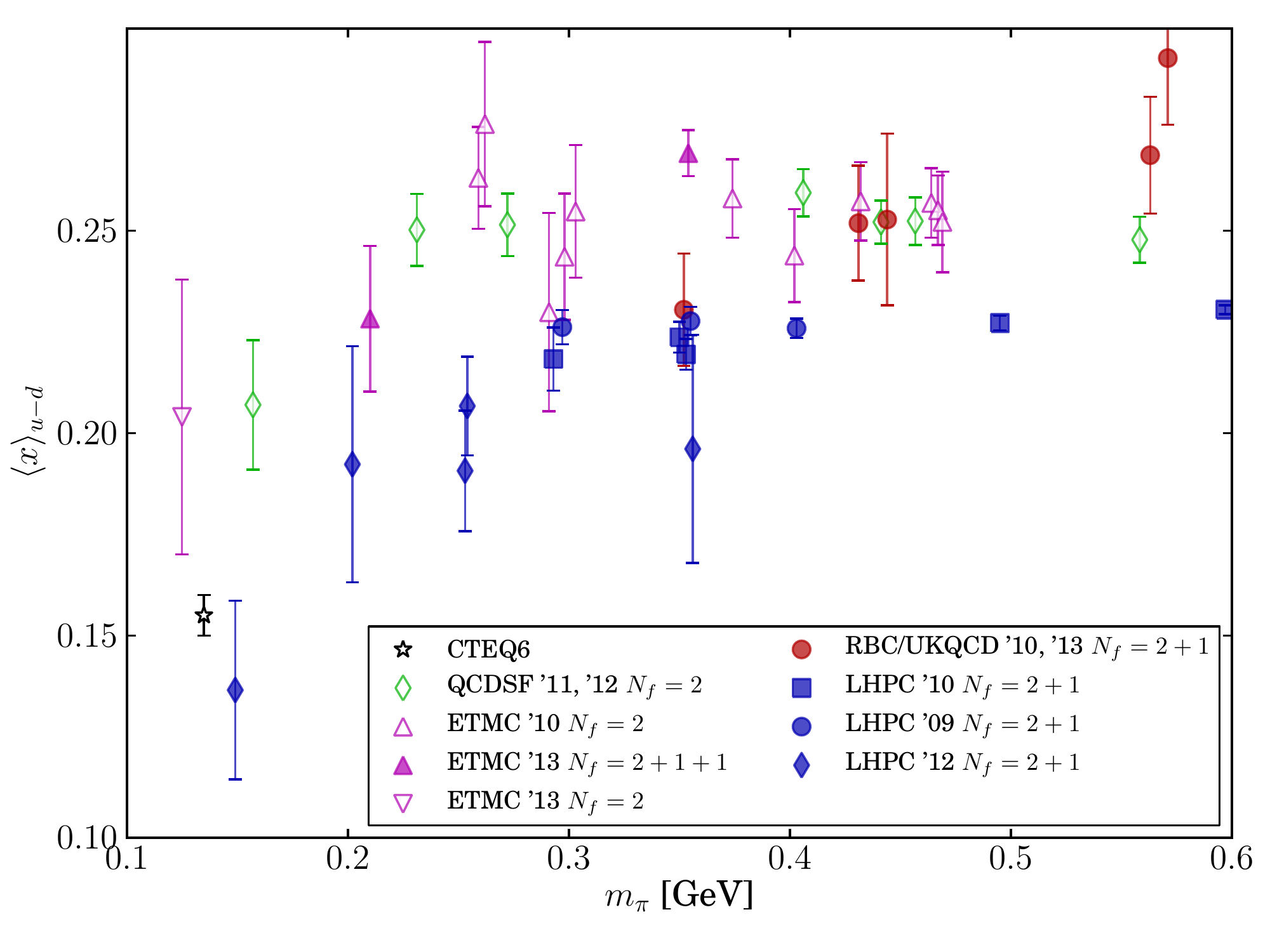}
    \caption{\label{fig:xv-summary}
      Summary of quark momentum fraction $\la x\ra_{u-d}^{\overline{MS}(2\text{ GeV})}$
      lattice results~\cite{Alexandrou:2013joa,Aoki:2010xg,Bali:2012av,Pleiter:2011gw,Bratt:2010jn,Green:2012ud}.
    }
  \end{minipage}~
  \hspace{.03\textwidth}~
  \begin{minipage}{.48\textwidth}
    \centering
    \includegraphics[width=\textwidth]{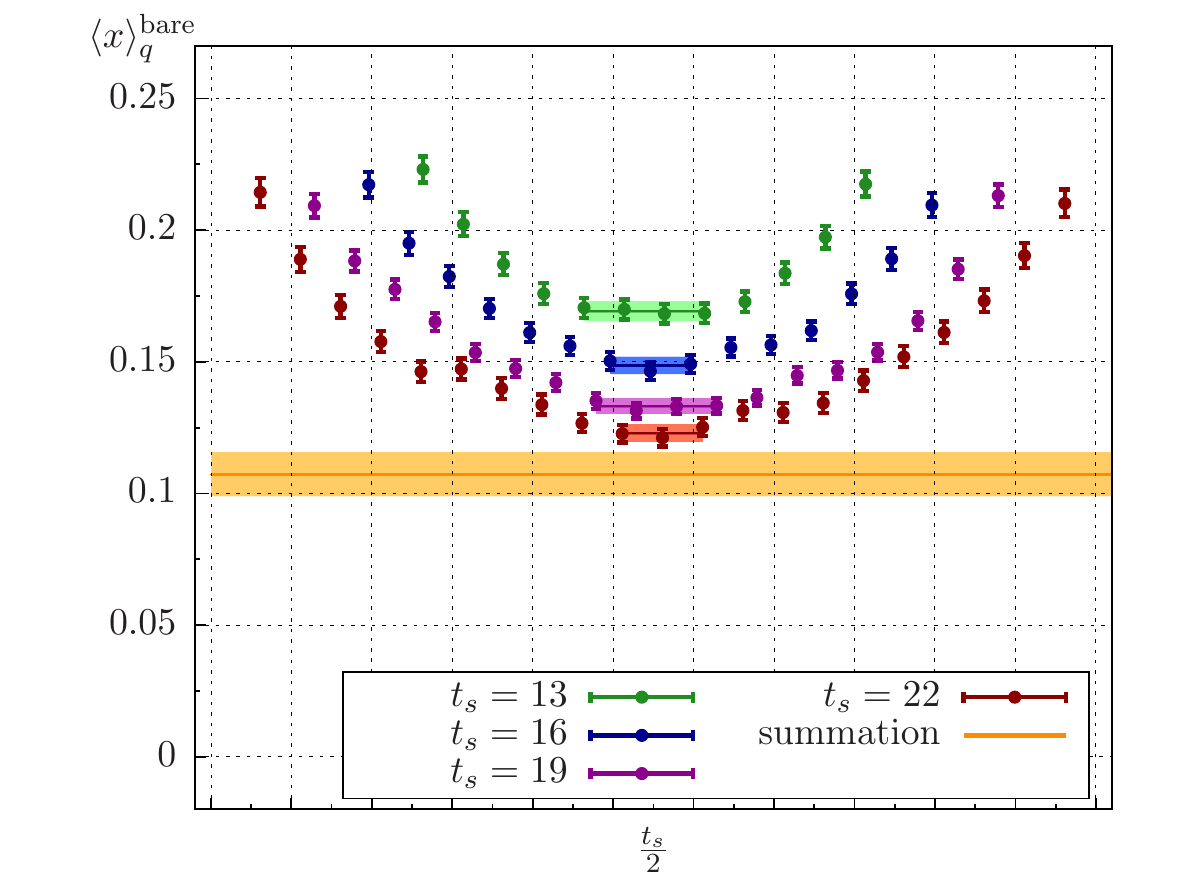}
    \caption{\label{fig:xv-exc-states}
      Excited state contributions to bare $\la x\ra_{u-d}$ and their removal using
      the summation method~\cite{Jager:2013kha}.
    }
  \end{minipage}
\end{figure}

Another ``benchmark'' quantity is the isovector quark momentum fraction.
It is measured in DIS experiments and, although its value depends on 
phenomenological models of parton distribution functions, different
parameterizations yield agreeing results.
Lattice calculation of this quantity involves the quark energy-momentum tensor operator,
\begin{equation}
\langle N(P)|\bar{q}\gamma_{\{\mu} \overset{\leftrightarrow}{D}_{\nu\}} q|N(P)\rangle 
  = \la x\ra_{q} \, \bar{u}_P \,P_{\{\mu}\gamma_{\nu\}} \, u_P\,,
\end{equation}
and typically is converted to $\overline{MS}(2\text{ GeV})$ using the RI/MOM 
method~\cite{Martinelli:1994ty}.
Until recently, lattice results overwhelmingly overestimated the phenomenological 
value by $40-60\%$.
Newer results closer to the physical pion mass tend to approach the experimental
value (Fig.~\ref{fig:xv-summary}).
Such behavior is in agreement with corrections computed in Chiral Perturbation Theory (ChPT),
$\delta^\text{ChPT}\la x\ra_{u-d}\sim m_\pi^2\log m_\pi^2$,
indicating that this quantity may change rapidly at lighter pion masses, 
thus precluding reliable chiral extrapolations.
Many recent studies point out that this quantity suffers from substantial 
excited state effects~\cite{Alexandrou:2011aa,Jager:2013kha,Bali:2013nla}, 
see Fig.~\ref{fig:xv-exc-states},
increasingly so towards the physical pion mass~\cite{Green:2012ud}, where subtraction 
of excited state contributions has lead to agreement with experiment.

\subsection{Nucleon radius and magnetic moment\label{sec:benchmk-radii-mag}}

\begin{figure}[ht!]
  \begin{minipage}{.495\textwidth}
    \centering
    \includegraphics[width=\textwidth]{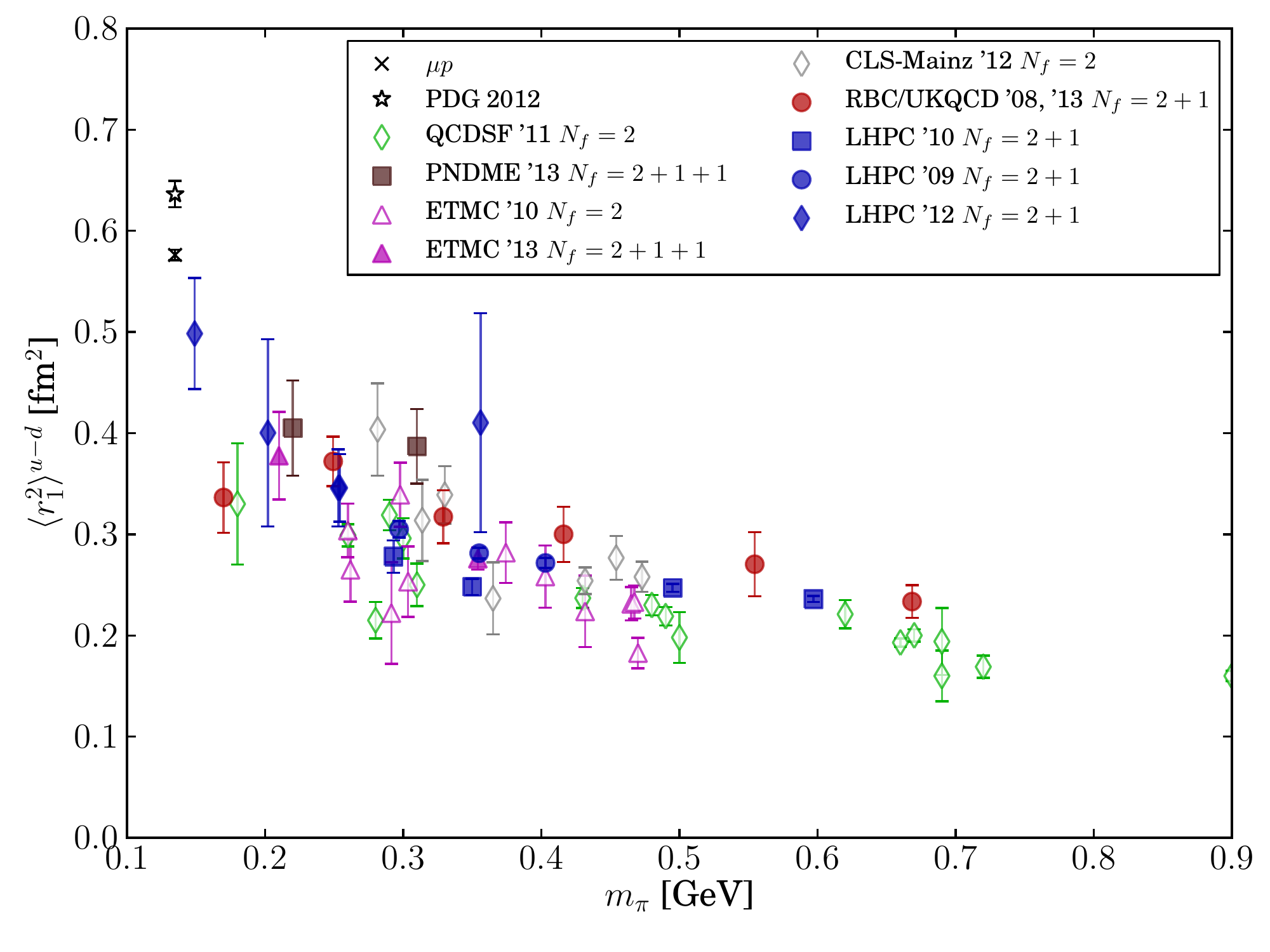}
    \caption{\label{fig:r1v-summary}
      Summary of $(r_1^2)^v$ lattice
results~\cite{Collins:2011mk,Bhattacharya:2013ehc,Capitani:2012ef,Alexandrou:2013joa,Lin:2013bxa,Syritsyn:2009np,Bratt:2010jn,Green:2012ud}.}
  \end{minipage}
  \begin{minipage}{.495\textwidth}
    \includegraphics[width=\textwidth]{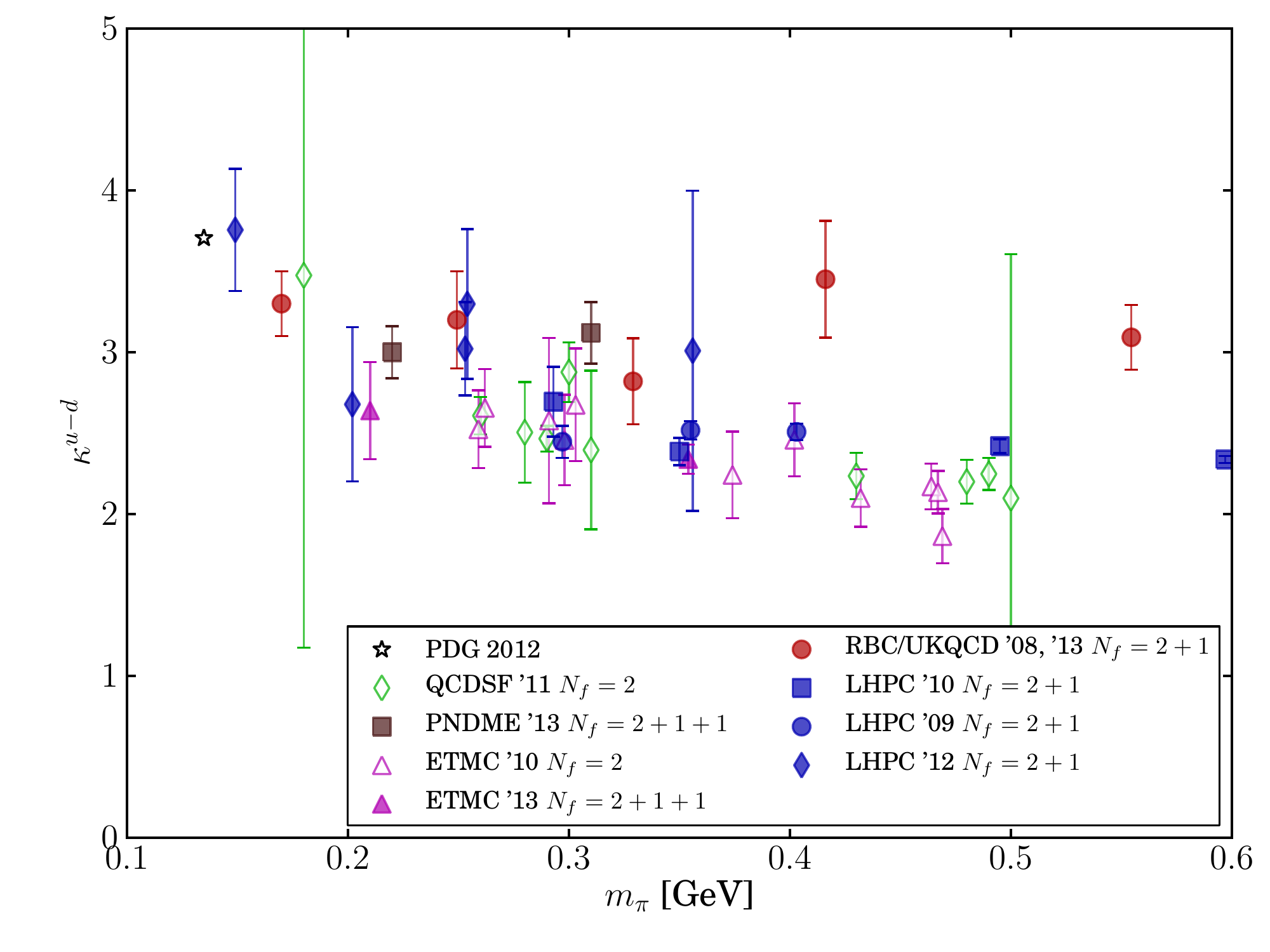}
    \caption{\label{fig:kappav-summary}
      Summary of $\kappa^v$ lattice
results~\cite{Collins:2011mk,Bhattacharya:2013ehc,Alexandrou:2013joa,Lin:2013bxa,Yamazaki:2009zq,Syritsyn:2009np,Bratt:2010jn,Green:2012ud}.}
  \end{minipage}
\end{figure}

Electromagnetic structure of the nucleon is characterized with 
the Dirac and Pauli form factors $F_{1,2}^q$:
\begin{equation}
\label{eqn:vector-ff}
\langle N(P + q)|\bar{q}\gamma^{\mu} q|N(P)\rangle 
  = \bar{u}_{P+q} \big[F_1^q(Q^2) \gamma^\mu +
    F_2^q(Q^2)\frac{i\sigma^{\mu\nu}q_\nu}{2 M_N}\big] u_P\,,
\quad Q^2 = -q^2\,,
\end{equation}
which will be discussed in more detail in Sect.~\ref{sec:formfac}.
The small-$Q^2$ behavior of these form factors,
$F^q(Q^2) = F(0)\big[1 + \frac16 Q^2 (r^2)^q +{\mathcal O}(Q^4) \big]$,
is characterized by Dirac and Pauli $(r_{1,2}^2)^q)$ radii of charge 
and (anomalous) magnetization distributions in the nucleon\footnote{
  These quantities should not be literally understood as radii because of 
  relativistic nucleon recoil taking place in measuring these form factors.
  Note also that experimentalists usually report electric and magnetic Sachs form factors 
  and the corresponding radii $(r_{E,M}^2)$~\cite{Beringer:1900zz}, 
  instead of $(r_{1,2})^2$; these pairs of quantities are linear combinations of
  each other.}.
Calculations of the isovector Dirac radius $(r_1^2)^v$ are tremendously 
important as a benchmark of lattice QCD, but even more so because of 
the persisting experimental discrepancy in the proton electric radius $(r^2_{Ep})$ 
between measurements involving electrons and muons~\cite{Pohl:2010zza}, 
which might be a signature of new physics phenomena.
The two experimental points in Fig.~\ref{fig:r1v-summary} demonstrate this
discrepancy in terms of $(r_1^2)^v$, together with a summary of results from
different lattice groups.
Similarly to $\langle x\rangle_{u-d}$, the isovector radius $(r_1^2)^v$ is also
strongly affected by low-energy QCD dynamics, diverging in the chiral limit as 
$\delta^\text{ChPT}(r_1^2)^v\sim\log m_\pi^2$,
a likely reason why calculations at heavier pion masses result in values
$\approx50\%$ below experiment.
However, many recent lattice QCD calculations with decreasing pion masses 
do not show a sufficient upward trend.
A calculation with $N_f=2+1$ dynamical ${\mathcal O}(a^2)$-improved Wilson fermions 
in range $m_\pi\approx250\rightarrow150\text{ MeV}$ demonstrated
that systematic effects from excited states increase dramatically, especially
below $m_\pi\lesssim200\text{ MeV}$, and their elimination with the simple 
``summation'' method is sufficient to achieve agreement with experiment~\cite{Green:2012ud}.
It is also encouraging that the statistical accuracy at the lightest pion mass
is comparable to the discrepancy between the two experiments
(Fig.~\ref{fig:r1v-summary}) and, if improved, may contribute to the resolution 
of the experimental controversy.
It is worth noting that ``radii'' are usually extracted from form factors using
phenomenological fits such as the dipole form.
In order to eliminate dependence on fit models, calculations at larger spatial 
volumes must be performed to have access to smaller values of $Q^2$.
Finite volume contributions to $G_E(Q^2)$ and $(r_E^2)$ have been computed 
in effective theory~\cite{Hall:2012yx} and are sizable,
$\delta (r_E^2)^v\big|_{m_\pi L=4}\approx0.03 (\text{fm})^2$, 
and thus must be studied as well.

A summary of anomalous magnetic moment $\kappa_v = F_2^{u-d}(0)$ calculations is
presented in Fig.~\ref{fig:kappav-summary}. 
This quantity has milder dependence on the pion mass, and its chiral extrapolations
and recent calculations close to the physical point are in good agreement
with experiment. 
To compute this quantity, one has to extrapolate the Pauli form factor with
$Q^2\to0$; increasing the lattice spatial size $L_s$ will reduce the systematic errors 
and provide another stringent test of lattice QCD.

\section{Hadron wave functions}

\begin{figure}[ht!]
  \begin{minipage}{.31\textwidth}
    \centering
    \includegraphics[angle=90,width=\textwidth]{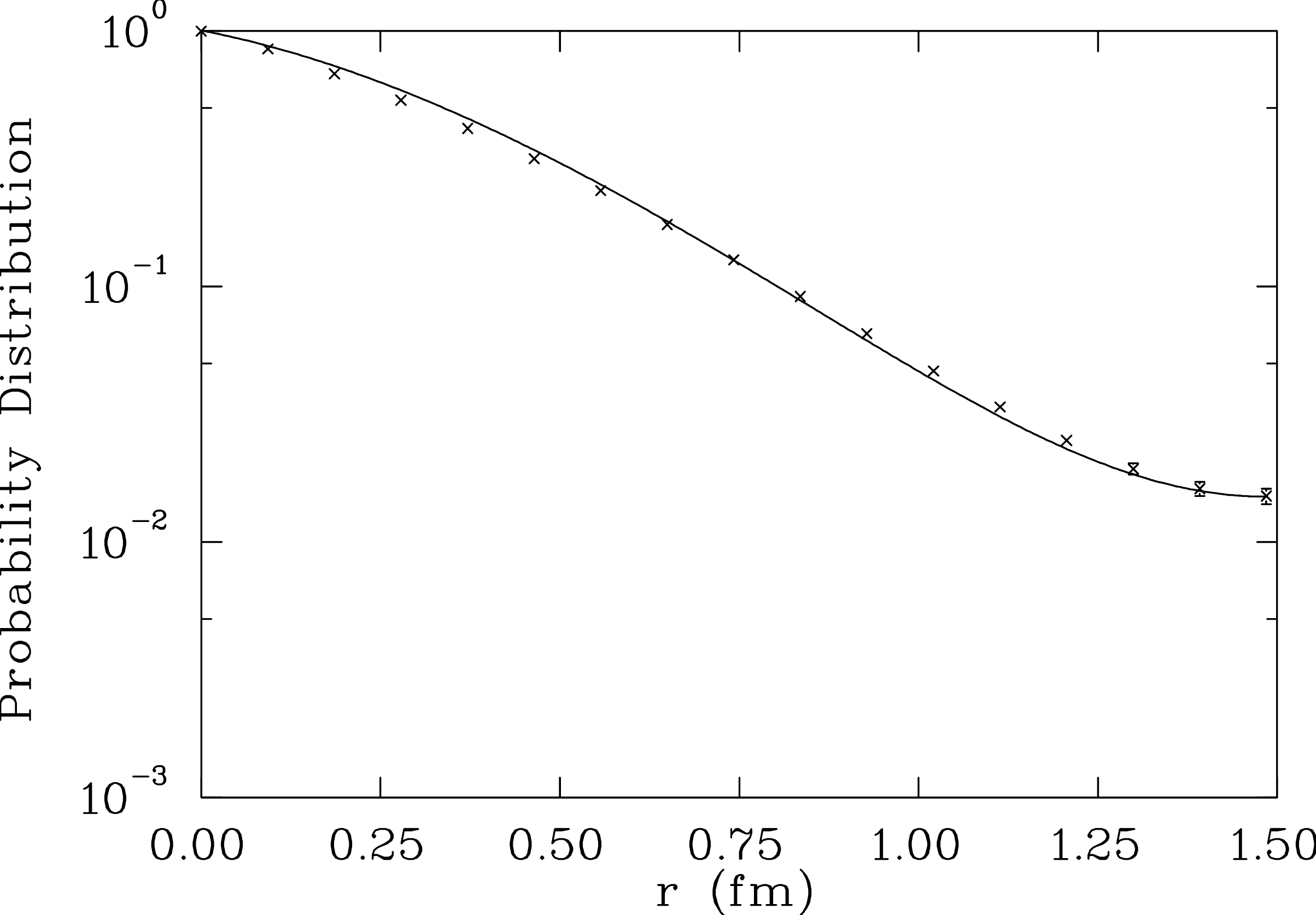}\\
    \includegraphics[width=\textwidth]{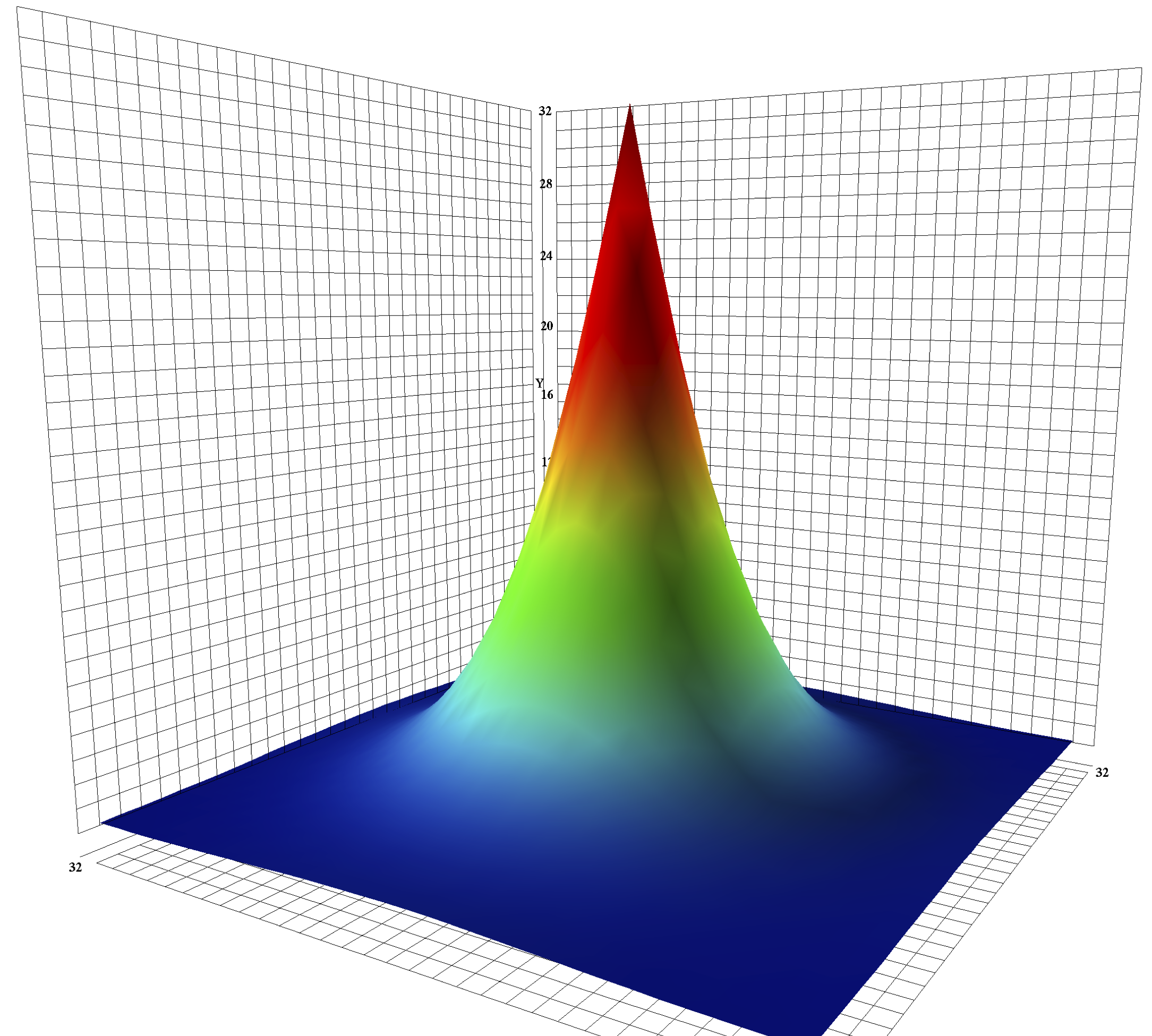}\\
  \end{minipage}~
  \hspace{.03\textwidth}~
  \begin{minipage}{.31\textwidth}
    \centering
    \includegraphics[angle=90,width=\textwidth]{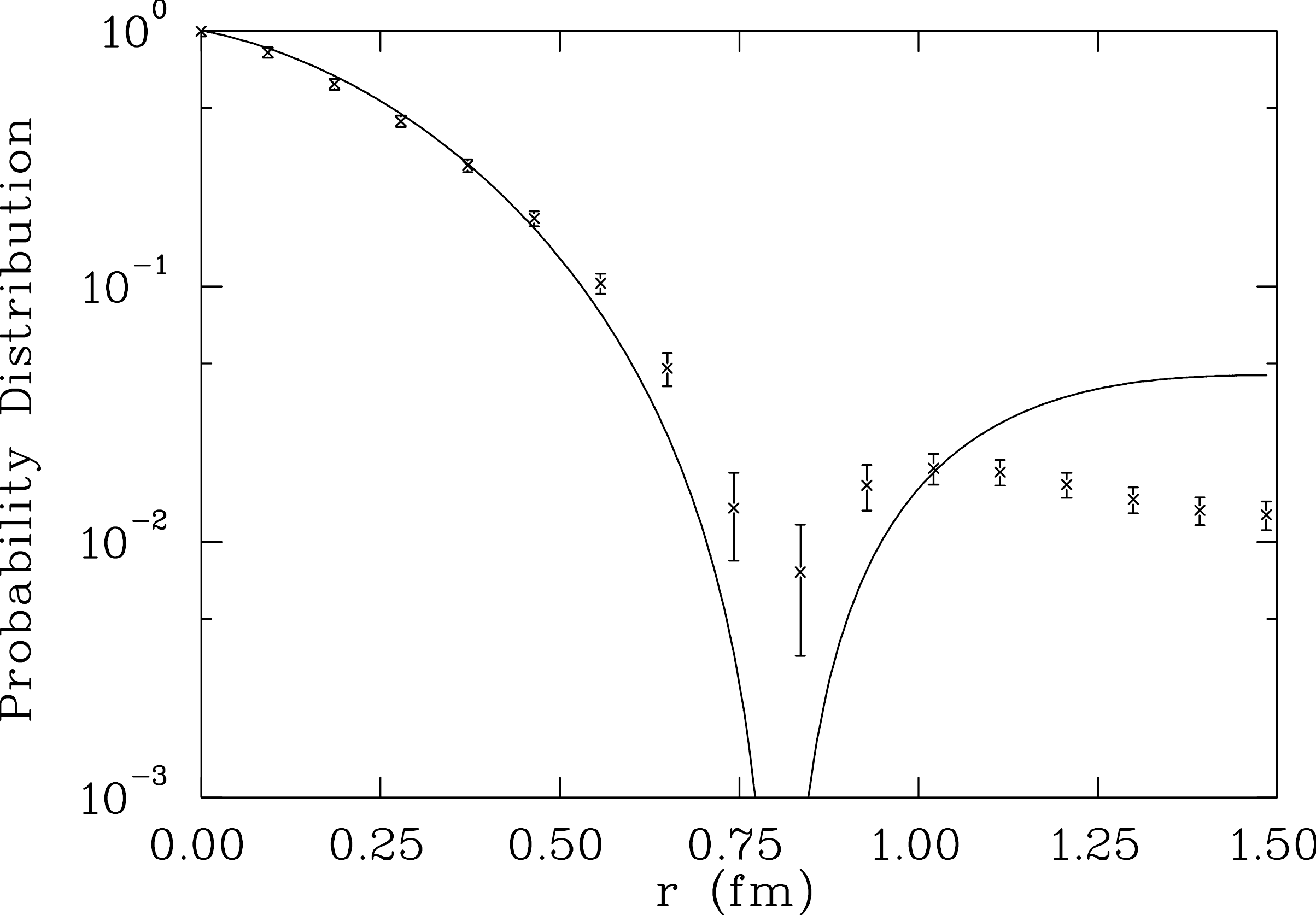}\\
    \includegraphics[width=\textwidth]{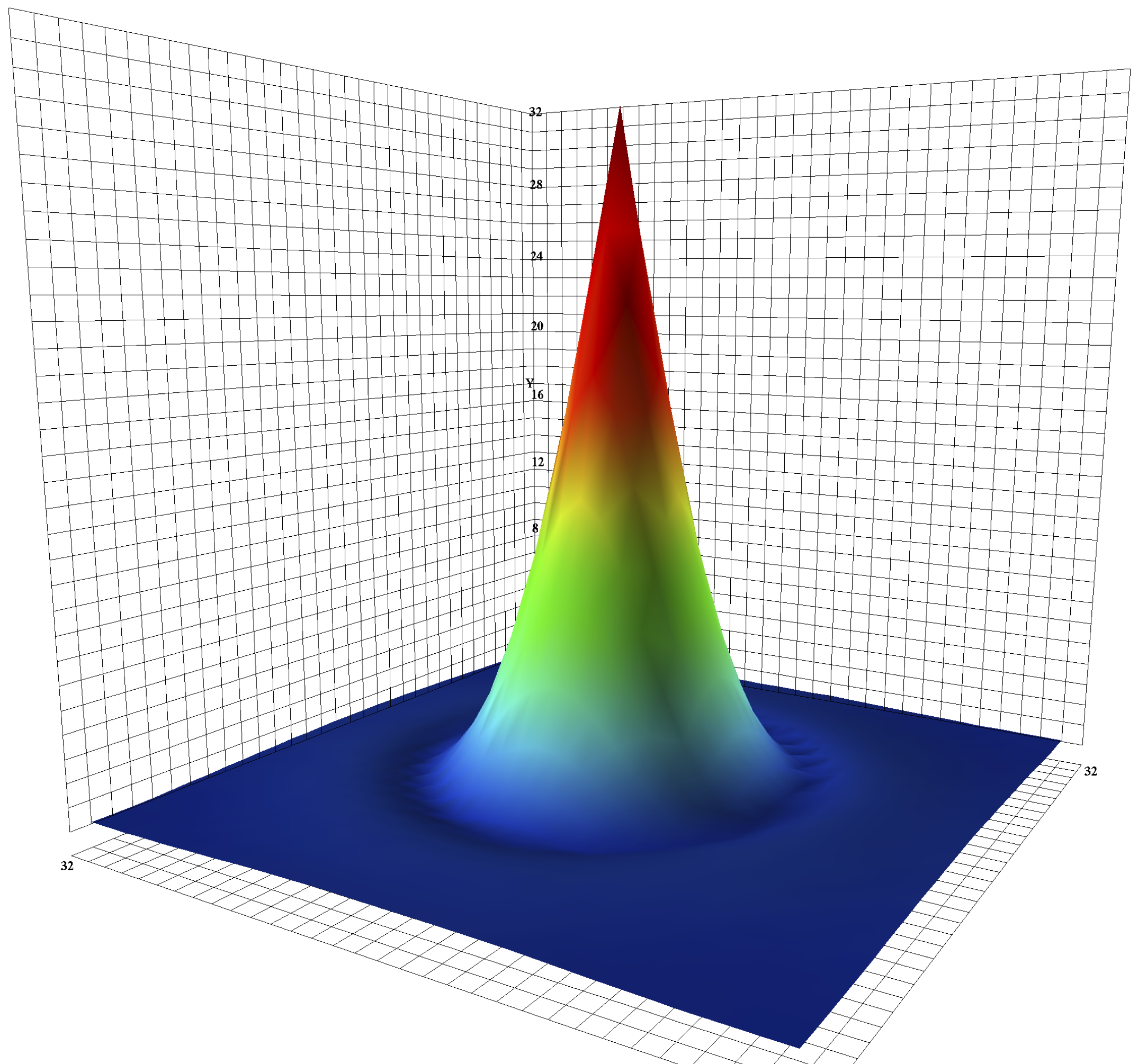}\\
  \end{minipage}~
  \hspace{.03\textwidth}~
  \begin{minipage}{.31\textwidth}
    \centering
    \includegraphics[angle=90,width=\textwidth]{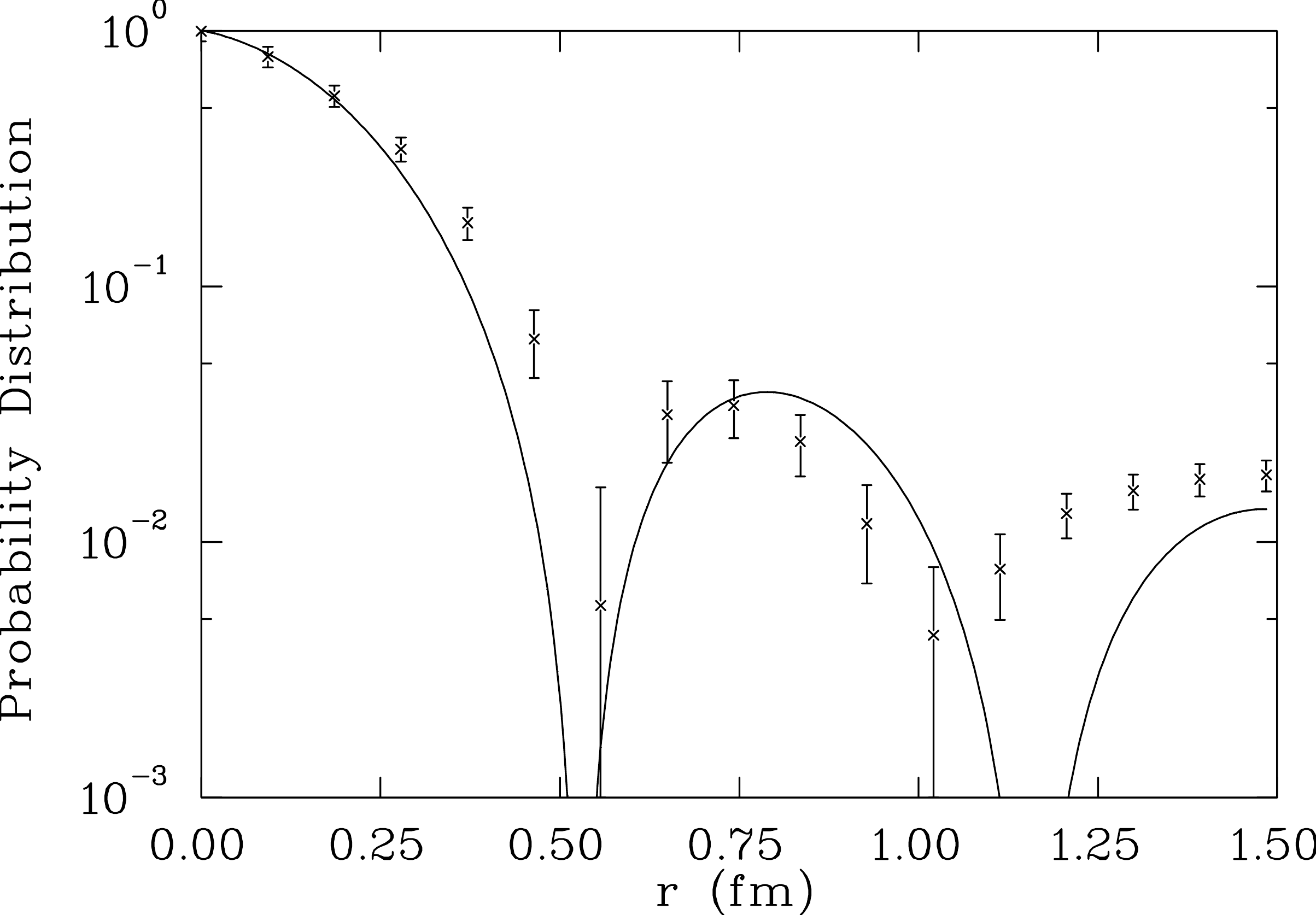}\\
    \includegraphics[width=\textwidth]{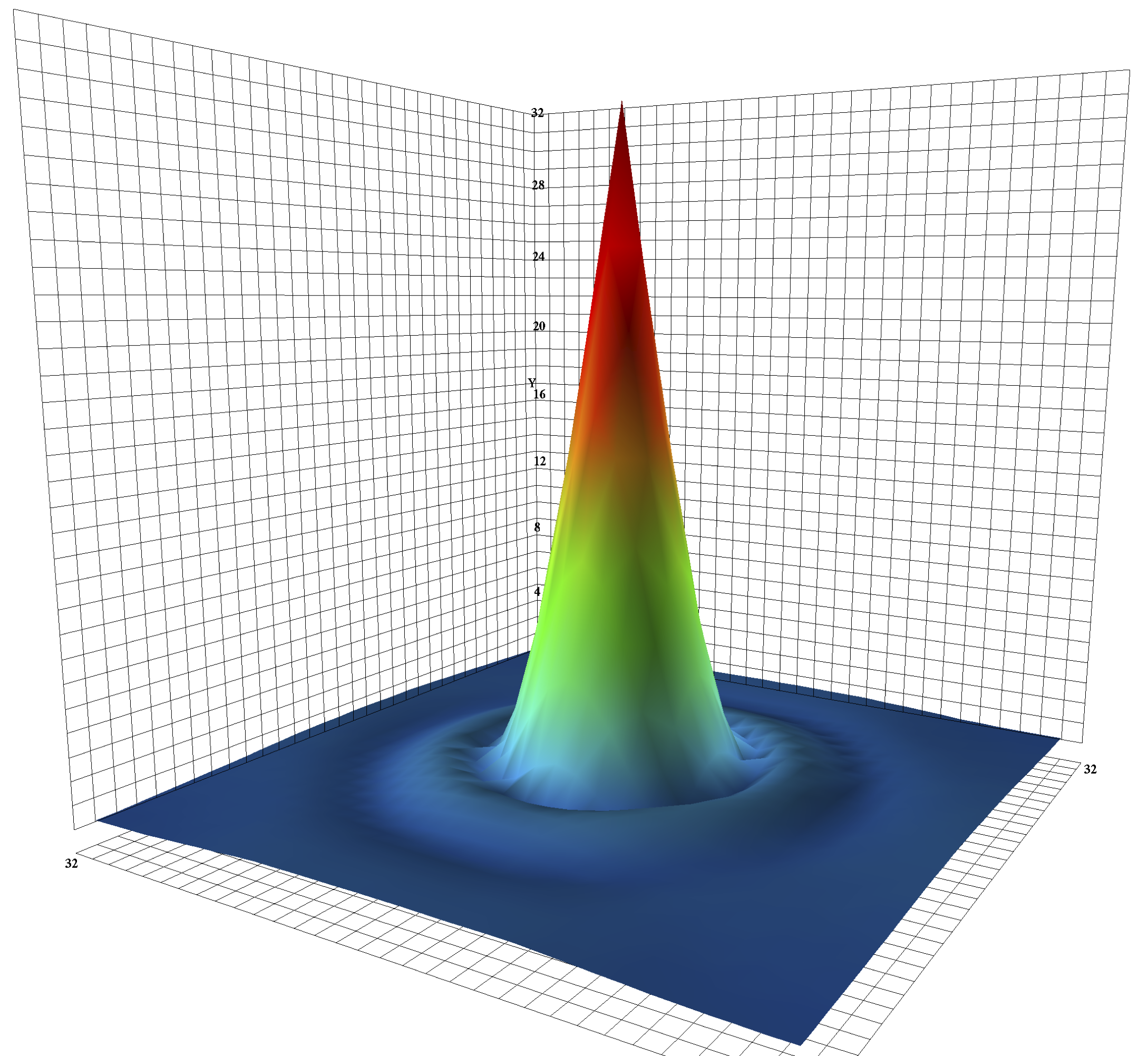}\\
  \end{minipage}~
  \caption{\label{fig:nucleon-wf}
    Wave functions of the nucleon and its radial excitations~\cite{Roberts:2013ipa}.
  }
\end{figure}

Lattice QCD provides a fascinating opportunity to study wave functions of quarks
in the nucleon and other hadrons.
Although such wave functions have only limited phenomenological meaning and 
are difficult to compare to experimental data, they can be very illuminating in our 
understanding of internal structure of hadrons and their excited states.
In a calculation very close to the physical point, radial profiles of quark density 
in the nucleon and its excited states have been studied~\cite{Roberts:2013ipa}, 
see Fig.~\ref{fig:nucleon-wf}.
Using a basis of four nucleon interpolating fields composed of quark sources with varied
smearing size, the CSSM collaboration has identified, in addition to the nucleon ground state, 
candidates for $n=1$ (Roper) and $n=2$ excitations, which have clearly visible nodes in
their radial quark density profile.

\begin{figure}[ht!]
  \includegraphics[width=\textwidth]{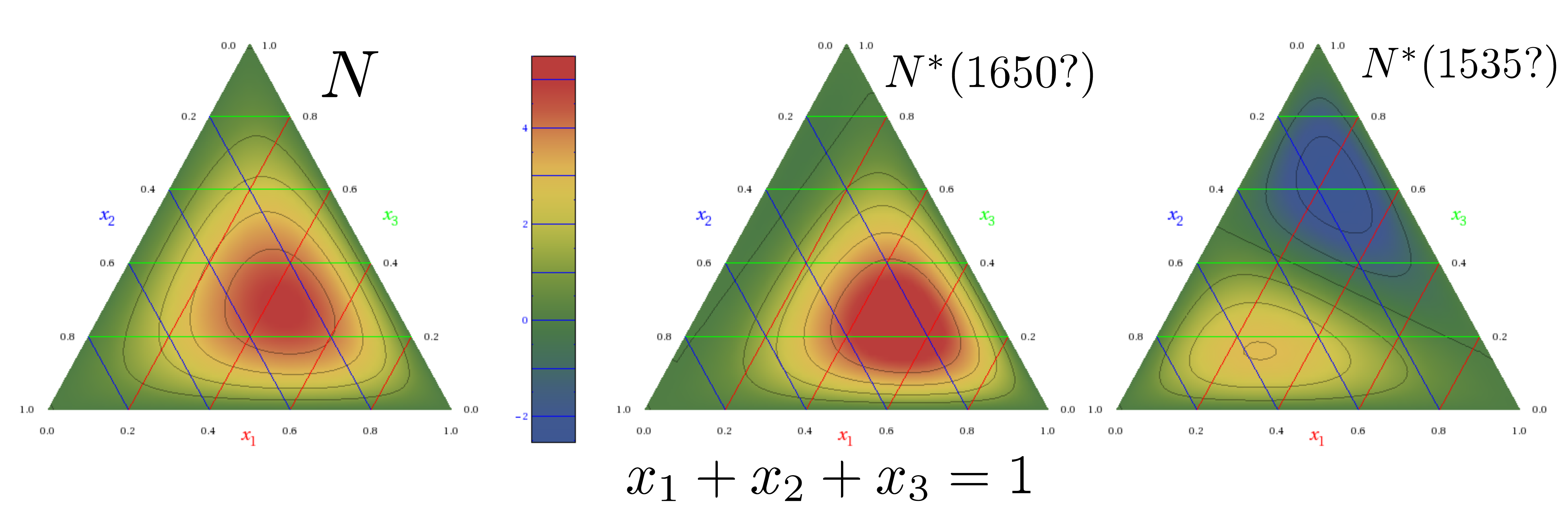}
  \caption{\label{fig:nucleon-da}
    Nucleon distribution amplitudes of valence quarks~\cite{Schiel:lat2013:2} in 
    coordinates $x_1+x_2+x_3=1$.}
\end{figure}

Hadron distribution amplitudes (DA), or partonic wave functions, describe 
hadron structure in terms of light-cone Fock states.
In case of the nucleon, at a sufficiently low scale these wave functions are dominated 
by three valence quarks carrying all of the boosted baryon momentum: $x_1 + x_2 + x_3=1$,
where $x_{1,2,3}$ are valence quark momentum fractions.
Distribution amplitudes cannot be computed directly on a lattice because probing a 
partonic wave function requires an operator with quarks separated along a light-cone direction.
Instead, DAs are parameterized as polynomials in $x_{1,2,3}$; the polynomial coefficients 
are called ``shape parameters'' and correspond to local operators calculable on a lattice.
Figure~\ref{fig:nucleon-da} shows results of a recent lattice calculation~\cite{Schiel:lat2013:2}
where distribution amplitudes of the nucleon and its excited states were computed using 
$N_f=2$ dynamical Wilson fermions.

\section{Hadron form factors}

\subsection{Nucleon electromagnetic (vector) form factors\label{sec:formfac}}

Nucleon Dirac and Pauli form factors $F_{1,2}(Q^2)$ defined in Eq.~(\ref{eqn:vector-ff}) 
are among the main characteristics of nucleon structure.
Interest in the nucleon form factors have been reignited in the recent years 
when more precise experiments became available in a wide range of momentum $Q^2$.
These form factors can also be considered benchmark quantities in addition to those 
discussed in Sec.~\ref{sec:benchmark}.
In addition to verifying the methodology, lattice QCD calculations of the form factors
may help resolve some experimental uncertainties. 
For example, proton form factor measurements are subject to corrections from $2\gamma$
exchange, while neutron studies must be conducted on nuclei and therefore have nuclear 
model uncertainties.

\begin{figure}[ht!]
  \begin{minipage}{.589\textwidth}
    \centering
    \includegraphics[width=\textwidth]{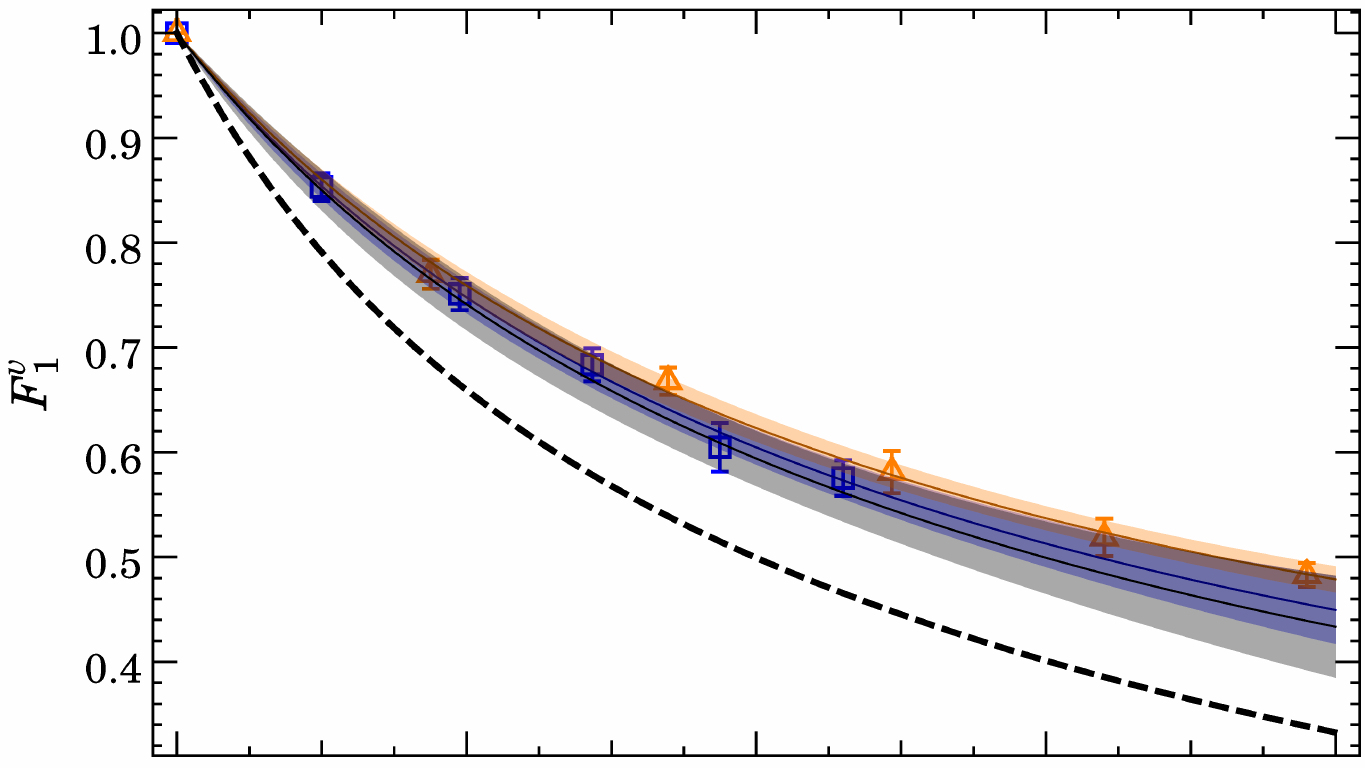}\\
    \includegraphics[width=\textwidth]{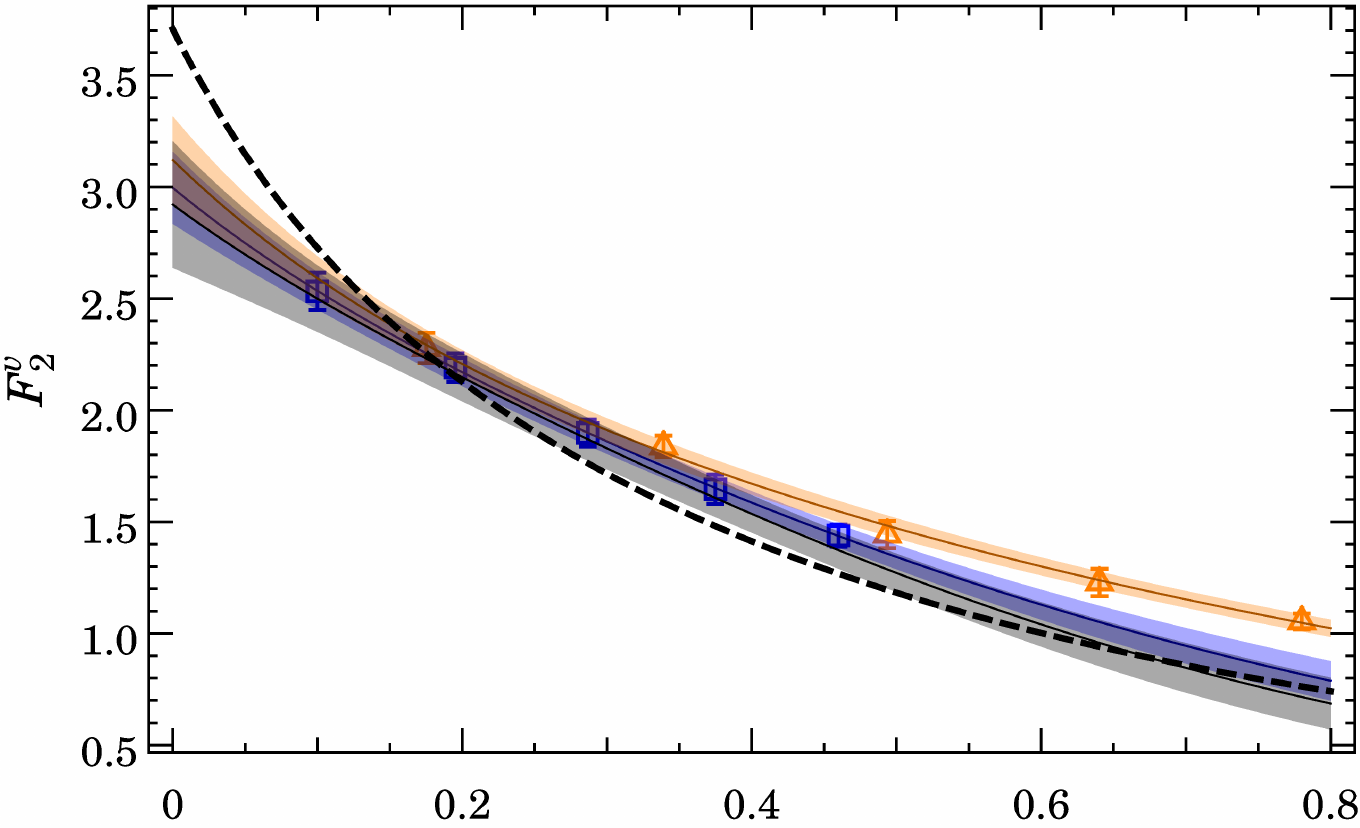}\\
    $Q^2\;[ \text{GeV}^2]$
  \end{minipage}
  \begin{minipage}{.401\textwidth}
    \centering
    \includegraphics[width=\textwidth]{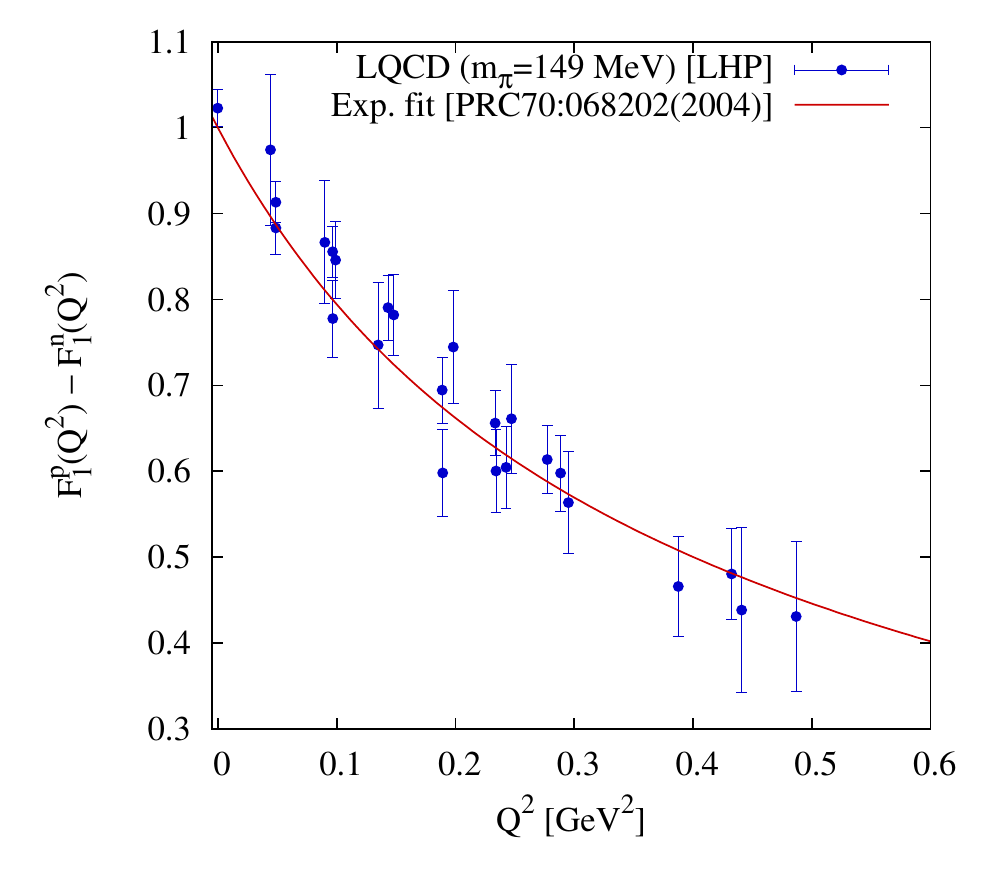}\\
    \vspace{-.05cm}
    \includegraphics[width=\textwidth]{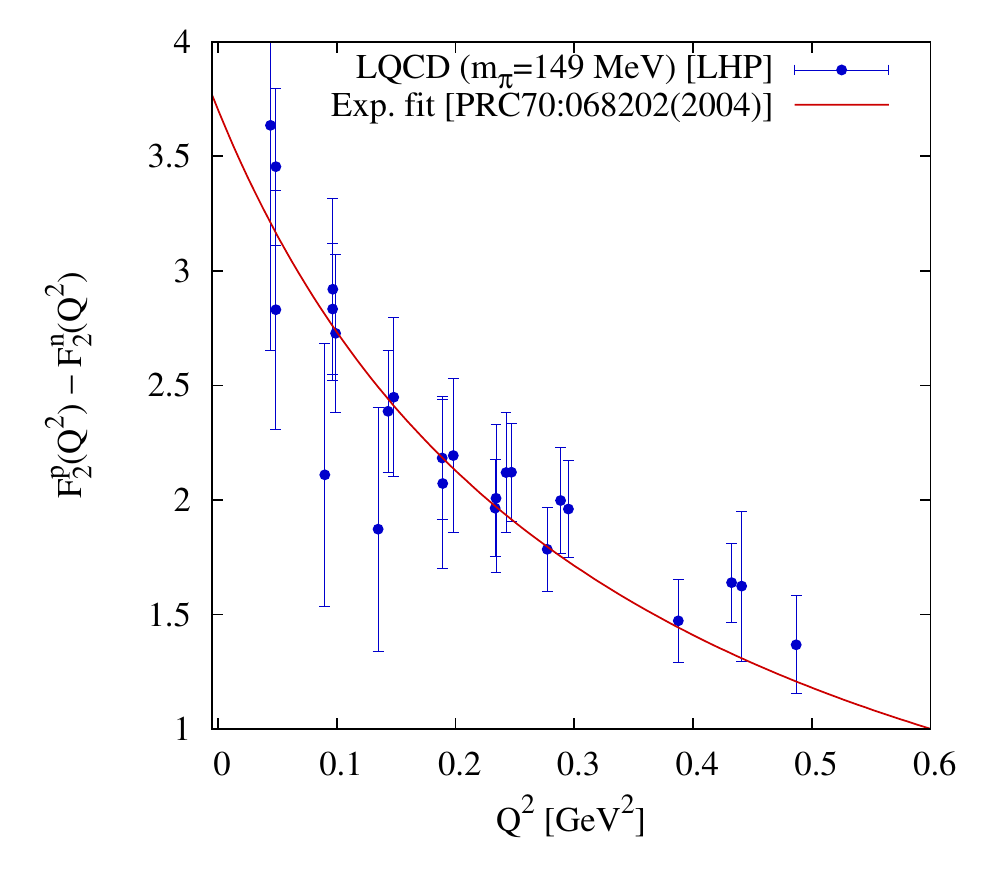}\\
    $Q^2\;[ \text{GeV}^2]$
  \end{minipage}
  \caption{\label{fig:emff-comp}
    Comparison of the nucleon isovector electromagnetic form factors $F_{1,2}(Q^2)$ computed with 
    $m_\pi=310,\,220\text{ MeV}$~\cite{Bhattacharya:2013ehc} 
    (left, triangles and squares, respectively) and
    $m_\pi=149\text{ MeV}$~\cite{Green:2013hja} (right). 
    Experimental parameterizations are from Ref.~\cite{Arrington:2007ux} (dashed line, left) 
    and Ref.~\cite{Kelly:2004hm} (solid line, right).}
\end{figure}

In Figure~\ref{fig:emff-comp}, two recent calculations~\cite{Green:2013hja,Bhattacharya:2013ehc} 
of nucleon isovector form factors $F_{1,2}^{u-d}(Q^2)$ are compared.
Both calculations use $N_f=2+1$ dynamic Wilson fermion action and 
incorporate advanced methods to isolate the ground state from excited states, 
which has been demonstrated to have dramatic impact on calculations of the isovector
radii~\cite{Green:2012ud}.
The panels on the right show form factors computed close to the physical point
($m_\pi=149\text{MeV}$), which agree nicely with the phenomenological fit to 
experimental data.
Calculations with heavier pion masses $m_\pi=310,\,220\text{ MeV}$ shown on the left
disagree with the experimental fit, especially form factor $F_1$ 
at intermediate momenta $Q^2\gtrsim0.5\text{ GeV}^2$.
This disagreement driven by heavy pion masses is surprising, since naively one would expect 
that the low-energy dynamics governed by the pion mass should not influence the structure
of the nucleon at momenta $Q^2\gg m_\pi^2$.
There is only little, borderline-significant downward trend in the form factor values 
between $m_\pi=310$ and $220\text{ MeV}$ in Fig.\ref{fig:emff-comp}(left), 
also noticed earlier in the range $m_\pi\approx300\ldots400\text{ MeV}$~\cite{Syritsyn:2009np}, 
suggesting that an abrupt change must take place between $m_\pi\approx200\text{ MeV}$ 
and the physical point.

\subsection{Nucleon axial form factors}

Nucleon axial form factors characterize nucleon structure with respect to the density 
of the axial vector quark current,
\begin{equation}
\label{eqn:axial-ff}
\langle N(P+q)|\bar{u}\gamma^{\mu}\gamma^5 u - \bar{d}\gamma^{\mu}\gamma^5 d|N(P)\rangle 
  = \bar{u}_{P+q} \big[G_A^q(Q^2) \gamma^\mu\gamma^5 +
    G_P^q(Q^2)\frac{\gamma^5 q^\mu}{2 M_N}\big] u_P\,,
\end{equation}
where $G_A$ and $G_P$ are axial and induced pseudoscalar form factors, respectively.
Experimental data on these form factors come from measurements of neutrino scattering, 
charged pion electroproduction and muon capture experiments~\cite{Bernard:2001rs}.

\begin{figure}[ht!]
  \begin{minipage}{.495\textwidth}
    \centering
    \includegraphics[width=\textwidth]{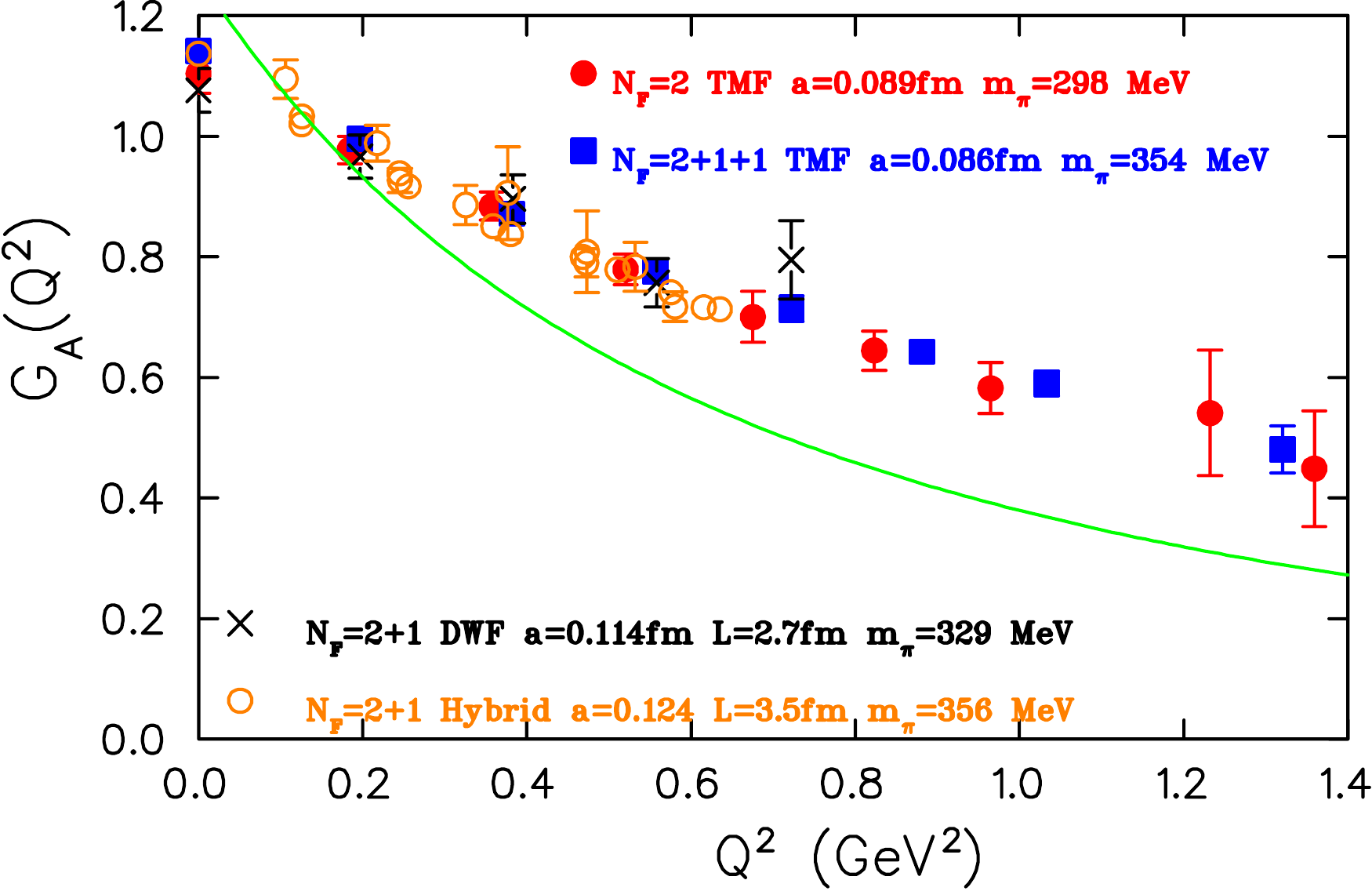}\\
  \end{minipage}
  \begin{minipage}{.495\textwidth}
    \centering
    \includegraphics[width=\textwidth]{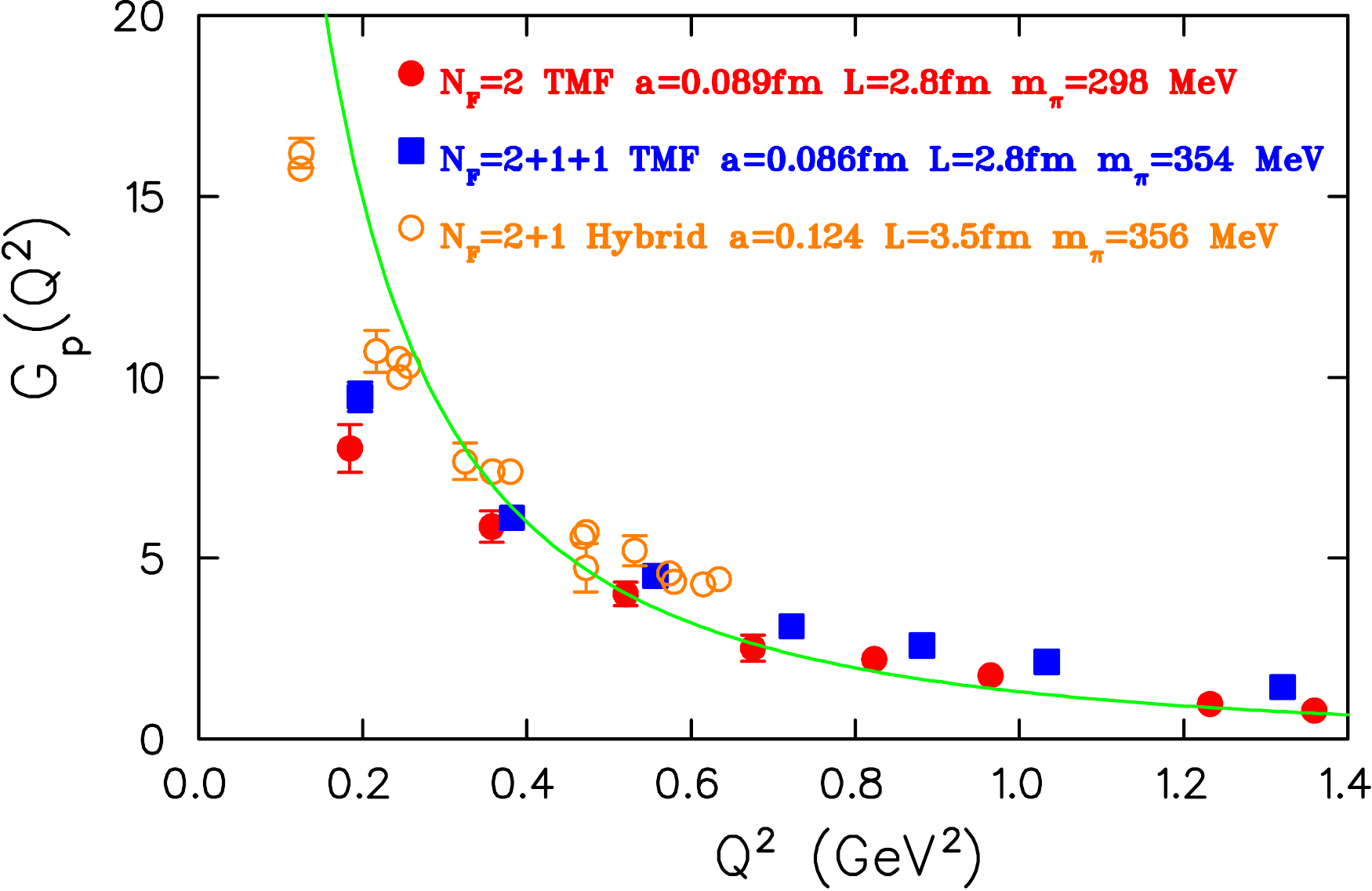}\\
  \end{minipage}
  \caption{\label{fig:ga-gp-summary} 
    Nucleon axial $G_A(Q^2)$ (left) and induced pseudoscalar $G_P(Q^2)$ 
    form factors~\cite{Alexandrou:2013joa}. 
    The solid lines are the dipole fit to $G_A$ (left) and the pion pole dominance fit to
    $G_P$ (right) experimental data (not shown).
  }
\end{figure}

Figure~\ref{fig:ga-gp-summary} displays the summary of lattice data on $G_{A,P}$ form factors 
calculated with various actions.
The left panel shows the axial form factor $G_A$ together with a phenomenological dipole 
fit to experimental points data (not shown).
The axial radius $r_A^2$ defined similarly to EM radii in Sec.~\ref{sec:benchmk-radii-mag} as
the slope $(-6)\big(dG_A(Q^2)/dQ^2\big)\big|_{Q^2=0}$
is underestimated by $\approx50\%$ for all pion mass values $m_\pi\ge213\text{ MeV}$, 
and it shows very little variation with the pion mass in the entire range 
$m_\pi=213\ldots373\text{ MeV}$~\cite{Alexandrou:2013joa}.
This non-trivial effect makes extrapolations based on baryon Chiral Perturbation Theory
questionable, since the NLO Lagrangian~\cite{Bernard:1998gv} does not contain any terms 
to account for $m_\pi$ dependence, thus requiring higher orders of baryon ChPT 
at pion masses as small as $m_\pi\approx200\text{ MeV}$.

The induced pseudoscalar form factor $G_P$ is special as it is governed by the intermediate
pion pole $\sim(Q^2+m_\pi^2)^{-1}$ in the coupling of the operator~(\ref{eqn:axial-ff}) 
to the nucleon.
Clearly, this form factor must depend strongly on the pion mass, becoming a steeper 
function of $Q^2$ with decreasing $m_\pi$.
However, the most recent $G_P$ results~\cite{Alexandrou:2013joa} exhibit 
very little dependence on $m_\pi$; this may be either a non-trivial low-energy 
effect similar to the one seen in $G_A$ form factor data, or, as suggested in 
Ref.~\cite{Alexandrou:2013joa}, may be caused by finite volume effects.
This fact makes accurate lattice calculation very challenging, especially in the region 
of low momenta $Q^2 \sim m_\mu^2$ relevant for phenomenology, where muon capture experiments
$\mu+p\to\nu_\mu + n$ can now measure $G_P$ with substantially improved 
precision~\cite{Andreev:2012fj}.

\subsection{Form factors of the pion and $\Lambda$}

\begin{figure}[ht!]
  \centering
  %\vspace{-.5cm}
  \begin{minipage}{.475\textwidth}
    \centering
    \includegraphics[height=.65\textwidth]{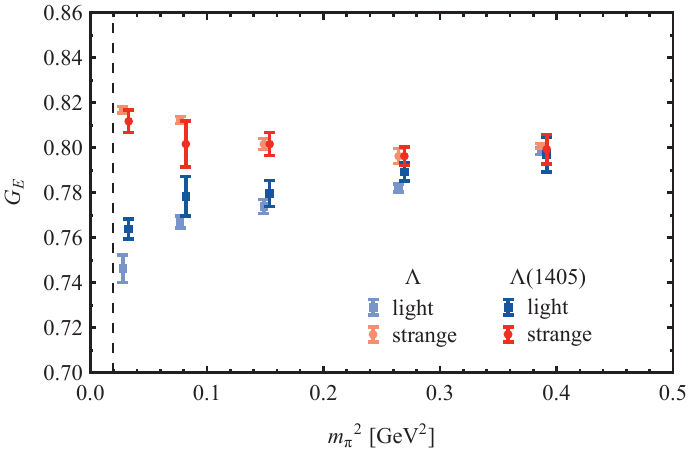}\\
    \caption{\label{fig:lambda1405-geff} 
      $G_E$ form factor at $Q^2=0.16\text{ GeV}^2$ for $\Lambda$ and
      $\Lambda(1405)$~\cite{Menadue:2013xqa}.
    }
  \end{minipage}~
  \begin{minipage}{.07\textwidth}
  \end{minipage}~
  \begin{minipage}{.475\textwidth}
    \centering
    \includegraphics[height=.65\textwidth]{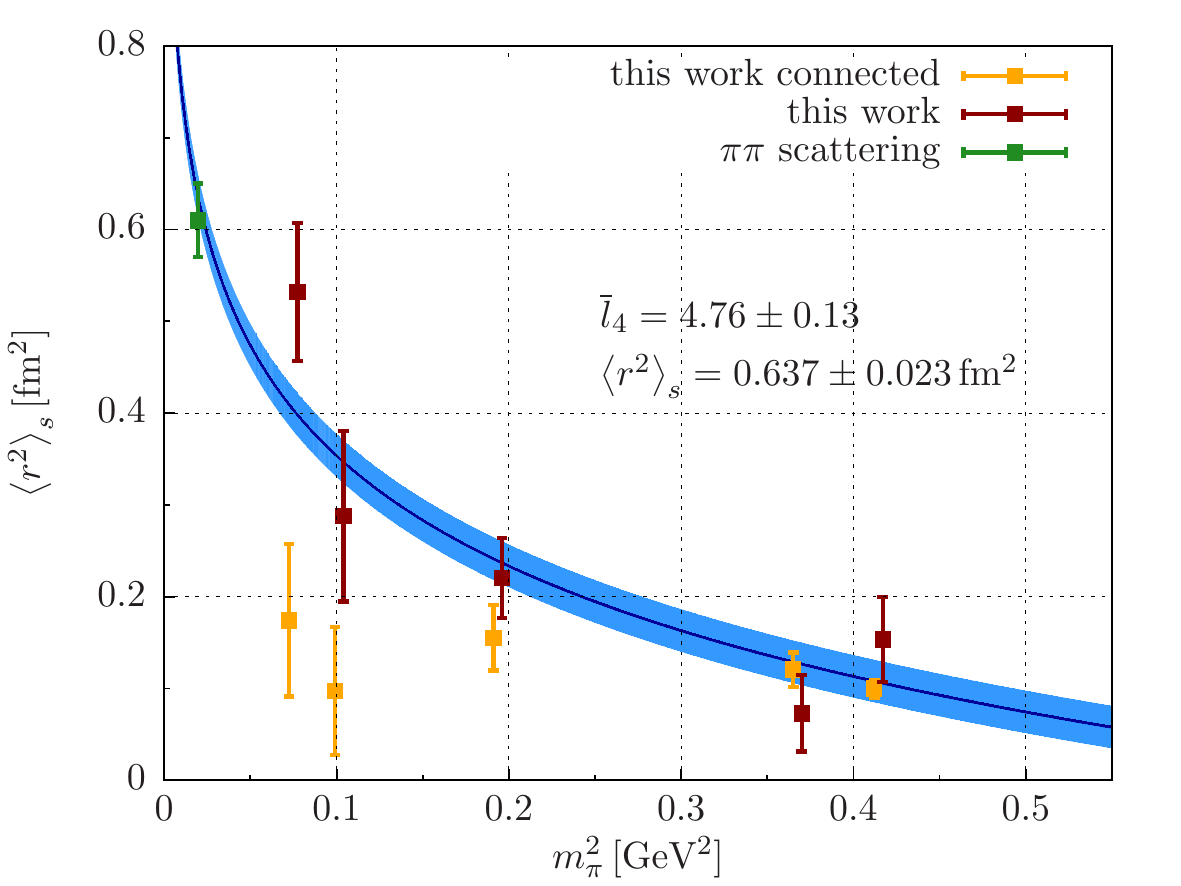}\\
    \caption{\label{fig:pion-scalar-rad}
      Scalar radius of the pion, full and connected-only terms~\cite{Gulpers:2013uca},
      $N_f=2$ dynamical fermions.
    }
  \end{minipage}
\end{figure}

Lattice QCD enables studying properties of hadrons that are 
not available or very difficult in experiments,
such as mesons or unstable baryons containing heavy quarks.
For example, computing structure of strange baryons enables testing hypotheses 
about their internal dynamics.
In a recent study~\cite{Menadue:2013xqa}, electric form factors of $\Lambda$ 
and $\Lambda(1405)$ were computed with the help of the variational method 
(see Fig.~\ref{fig:lambda1405-geff}).
From the picture, one can conclude that the mean squared radius of 
$s$-quark distribution is enhanced when comparing $\Lambda(1405)$ with $\Lambda$,
while the radius of light $u,d$-quarks is shrunk, as $m_\pi$ goes to the physical point.
Such differences support the theory that $\Lambda(1405)$, a $0({\frac12}^-)$ state,
has a significant component of the ``molecular'' $\overline{K}N$ state, in which the heavier
nucleon is surrounded by the cloud of the $\overline{K}=s\bar{q}_\text{light}$ meson.

Another example is the scalar radius of the pion for which only phenomenological estimates exist.
A few studies have been performed that disagreed with this estimate by a factor of two.
A recent study~\cite{Gulpers:2013uca}, however, has taken into account disconnected 
quark contractions, that turned out to be comparable in magnitude to the connected terms, 
see Fig.~\ref{fig:pion-scalar-rad}. 
The resulting values agree nicely with the NLO prediction of ChPT in a range of pion masses,
and the extrapolated value agrees with phenomenology at the physical pion mass.
This agreement confirms the validity of ChPT in the meson sector and provides 
an additional method to determine low energy constants of the theory, 
with the ultimate goal to determine all the parameters from first principles.

\section{Origin of the proton spin}

The proton spin puzzle is the experimental fact that the spins of quarks comprise
only $\approx30-50\%$ of the full $\frac12$-spin of the nucleon.
The missing part of the nucleon spin can be generated by the orbital motion of quarks and
angular momentum of the glue.
Size of their contributions can be computed on a lattice with the help of generalized form
factor formalism\footnote{
  Recently it has also been suggested that parton distribution functions can be studied
  directly on a lattice~\cite{Ji:2013dva}. }, 
in which quark and gluon momentum and angular momentum can be defined 
in a gauge-invariantly~\cite{Ji:1996ek}:
\begin{gather}
\la N(P+\frac12q) |\, T^{q,g}_{(\mu\nu)} \,| N(P-\frac12q)\ra
  = \bar{u}_{P+\frac12q} \big[
      A_{20}^{q,g} \gamma_{(\mu} P_{\nu)}
      + B_{20}^{q,g} \frac{P_{(\mu}\,i\sigma_{\nu)\rho} q^\rho}{2 M_N}
      + C_2^{q,g} \frac{q_{(\mu} q_{\nu)}}{M_N}
    \big] u_{P-\frac12q}
\end{gather}
where $T^{q,g}_{\mu\nu}$ is the energy-momentum tensor of quarks or gluons 
and $(\mu\nu)$ above denotes symmetrization over the indices and subtraction of the trace. 
The generalized form factors $\{A_{20},B_{20},C_2\}^{q,g}(Q^2)$ depend 
on the momentum transfer $Q^2=-q^2$ and their forward values at $Q^2\to0$ 
give the size of momentum fraction and angular momentum carried by quarks and gluons,
\begin{equation}
\la x \ra_{q,g} = A_{20}^{q,g}(0)\,,
\quad\quad
J_{q,g} = \frac12\big[A_{20}^{q,g}(0) + B_{20}^{q,g}(0)\big]\,.
\end{equation}
In the case of quarks, the angular momentum above is a sum of orbital momentum and spin;
for gluons, spin and orbital motion cannot be separated in a gauge-invariant way 
in this formalism.

\begin{figure}[ht!]
  \begin{minipage}{.495\textwidth}
    \centering
    \includegraphics[width=\textwidth]{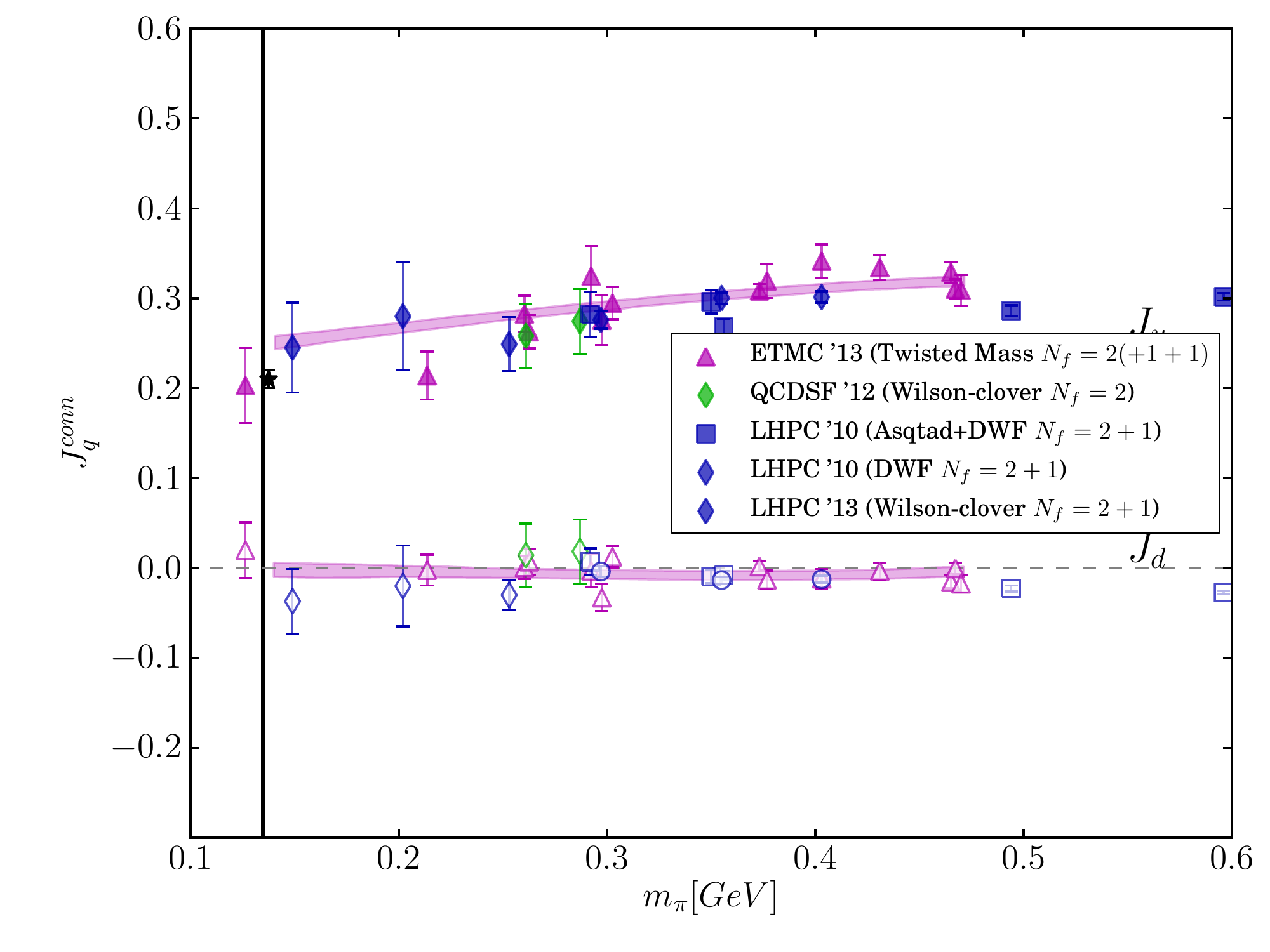}\\
  \end{minipage}
  \begin{minipage}{.495\textwidth}
    \centering
    \includegraphics[width=\textwidth]{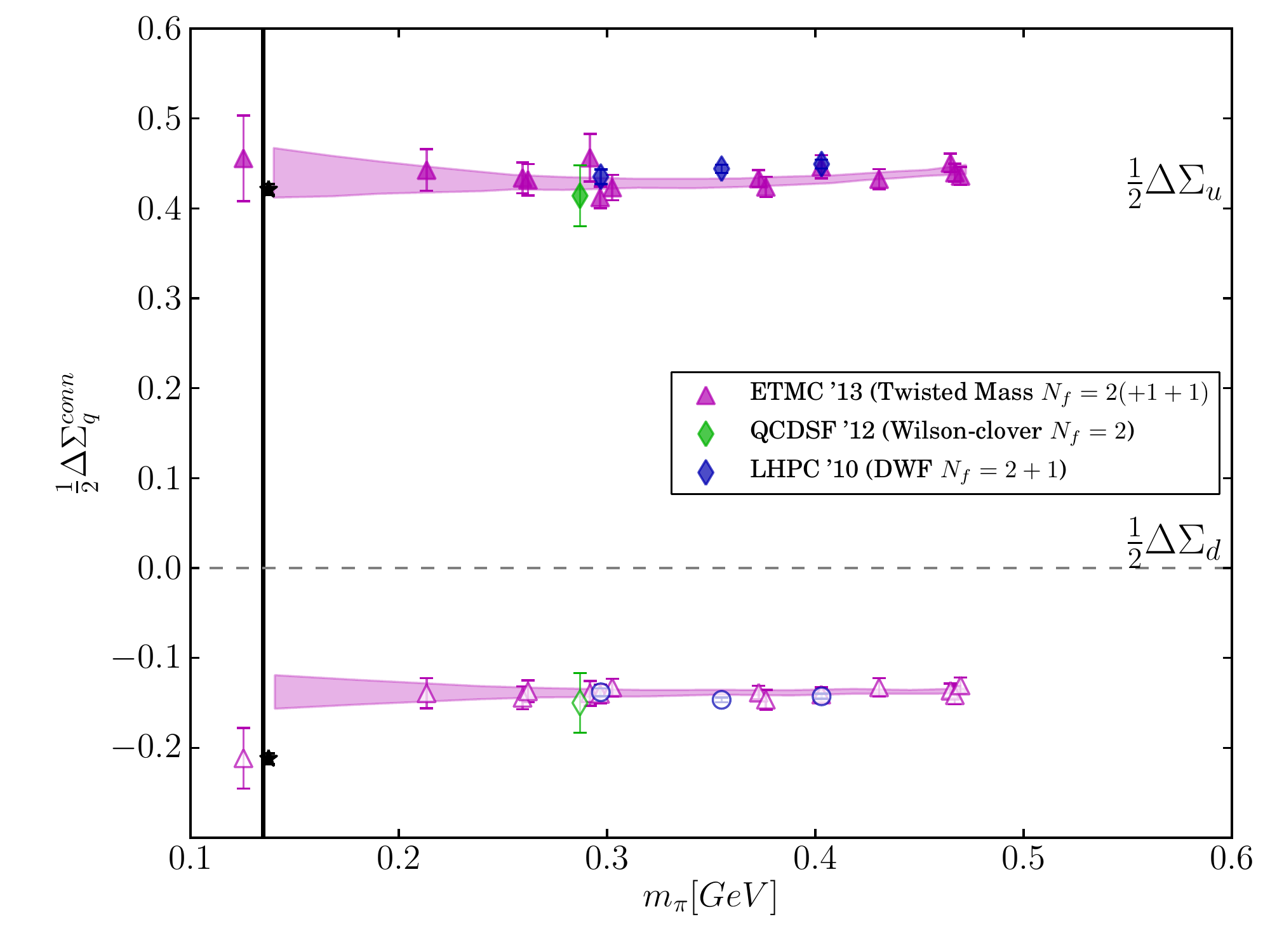}\\
  \end{minipage}
  \caption{\label{fig:lightq-angmom-spin} 
    Angular momentum $J_q$ (left) and spin $\frac12\Sigma_q$ 
    (right)~\cite{Alexandrou:2013joa,Sternbeck:2012rw,Bratt:2010jn,Syritsyn:2011vk} 
    of light quarks in the proton (connected contributions only) 
    computed with dynamical fermions.
  }
\end{figure}

\begin{figure}[ht!]
  \begin{minipage}{.495\textwidth}
    \centering
    \includegraphics[angle=-90,width=\textwidth]{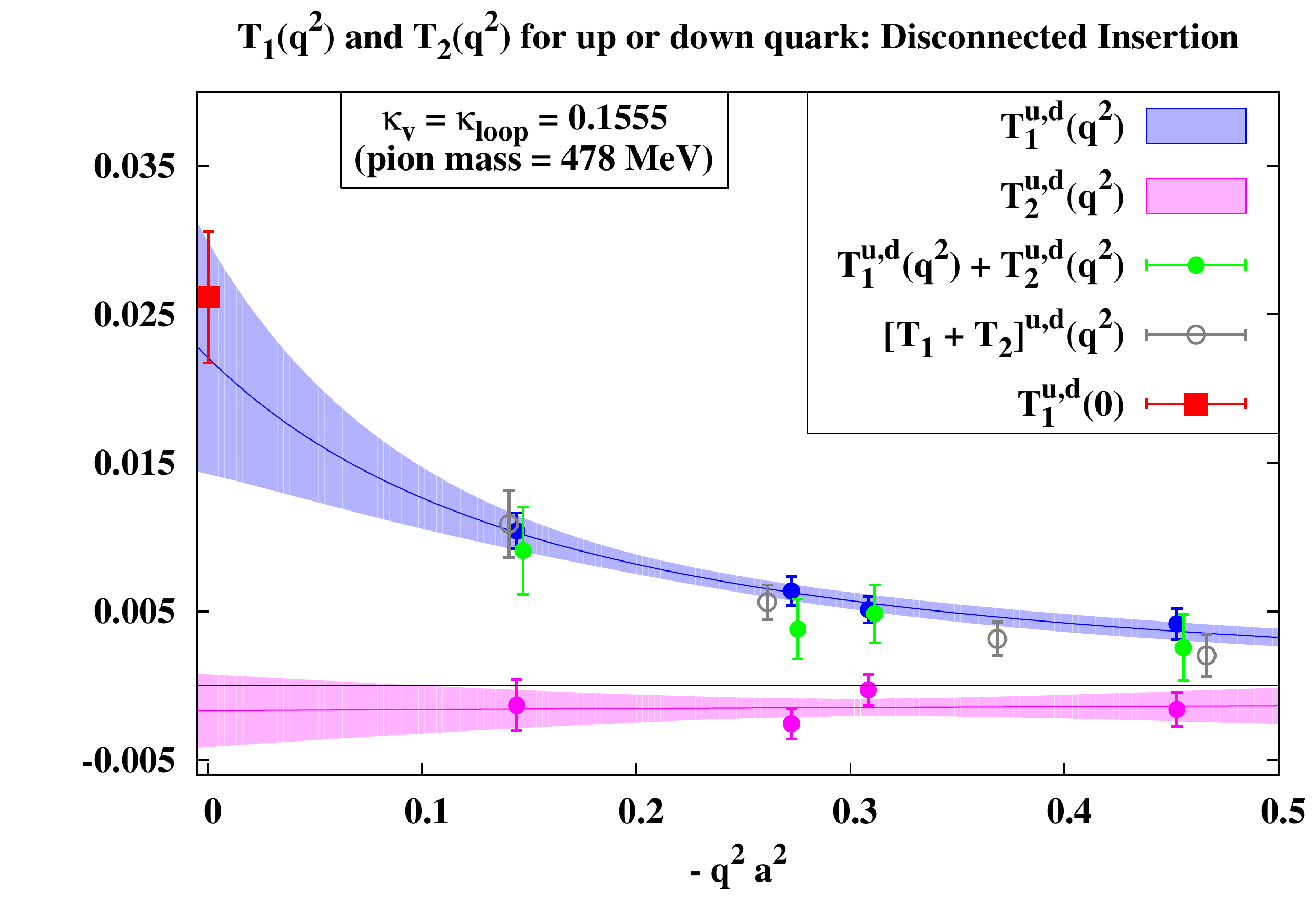}\\
  \end{minipage}
  \begin{minipage}{.495\textwidth}
    \centering
    \includegraphics[angle=-90,width=\textwidth]{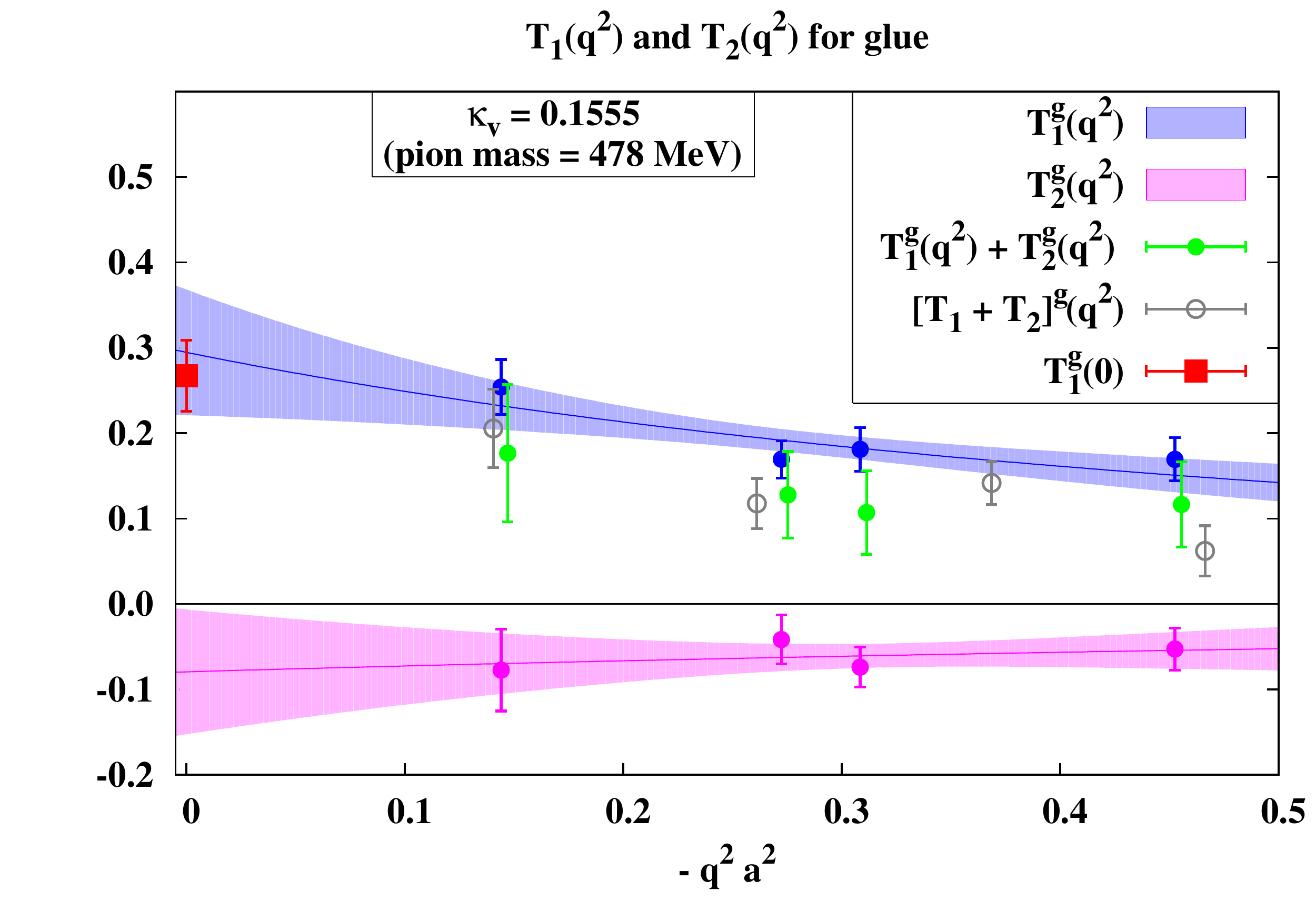}\\
  \end{minipage}
  \caption{\label{fig:discq-glue-angmom} 
    Generalized form factors $T_1\equiv A_{20}$, $T_2\equiv B_{20}$ and 
    angular momentum $2 J_q = T_1^q(0) + T_2^q(0)\equiv A_{20}^q(0) + B_{20}^q(0)$
    of quarks (disconnected contractions) (left) and 
    gluons (right) computed with quenched Wilson fermions~\cite{Deka:2013zha}.
  }
\end{figure}

Multiple groups have computed connected contributions to the light quark 
angular momentum $J^q$, see Fig.~\ref{fig:lightq-angmom-spin}(left) for the summary.
These quantities apparently have only mild dependence on the pion mass, and there
is good agreement across calculations done with a variety of different actions and lattice 
volumes.
Qualitatively, the angular momentum is carried only by the $u$-quark, 
and the $d$-quark angular momentum is dramatically smaller\footnote{
  The quark angular momentum is a scale-dependent quantity and the qualitative statements must
  be made with respect to a certain renormalization scheme and scale.
  In addition, isoscalar light-quark (angular) momentum mixes with that of the gluon.
  All results discussed in this review are converted to $\overline{MS}(2\text{ GeV})$
  and mixing with gluons are ignored.}.
The smallness of $J^d$, however, is not a trivial fact since the $d$-quark spin and orbital
angular momentum (OAM) appear to cancel each other, at least their connected parts (see
Fig.~\ref{fig:lightq-angmom-spin}, left).

The full calculation of all the contributions to the proton spin~\cite{Deka:2013zha} 
has been performed only with quenched fermions and relatively heavy pions
$m_\pi\ge478\text{ MeV}$ so far.
This simplification is justified in order to have a complete picture of 
separate contributions to the proton spin, the most challenging of which are 
the quark-disconnected contributions to $J^q$ and the gluon angular momenta $J^g$,
both shown in Fig.~\ref{fig:discq-glue-angmom}.
The disconnected light-quark angular momentum $J^{u+d}_\text{disc}$ is small 
(approximately 7\% of the proton spin), while the glue angular momentum was found 
to comprise $\approx25\%$ of the proton spin.

\begin{figure}[ht!]
  \begin{minipage}{.495\textwidth}
    \centering
    \includegraphics[width=\textwidth]{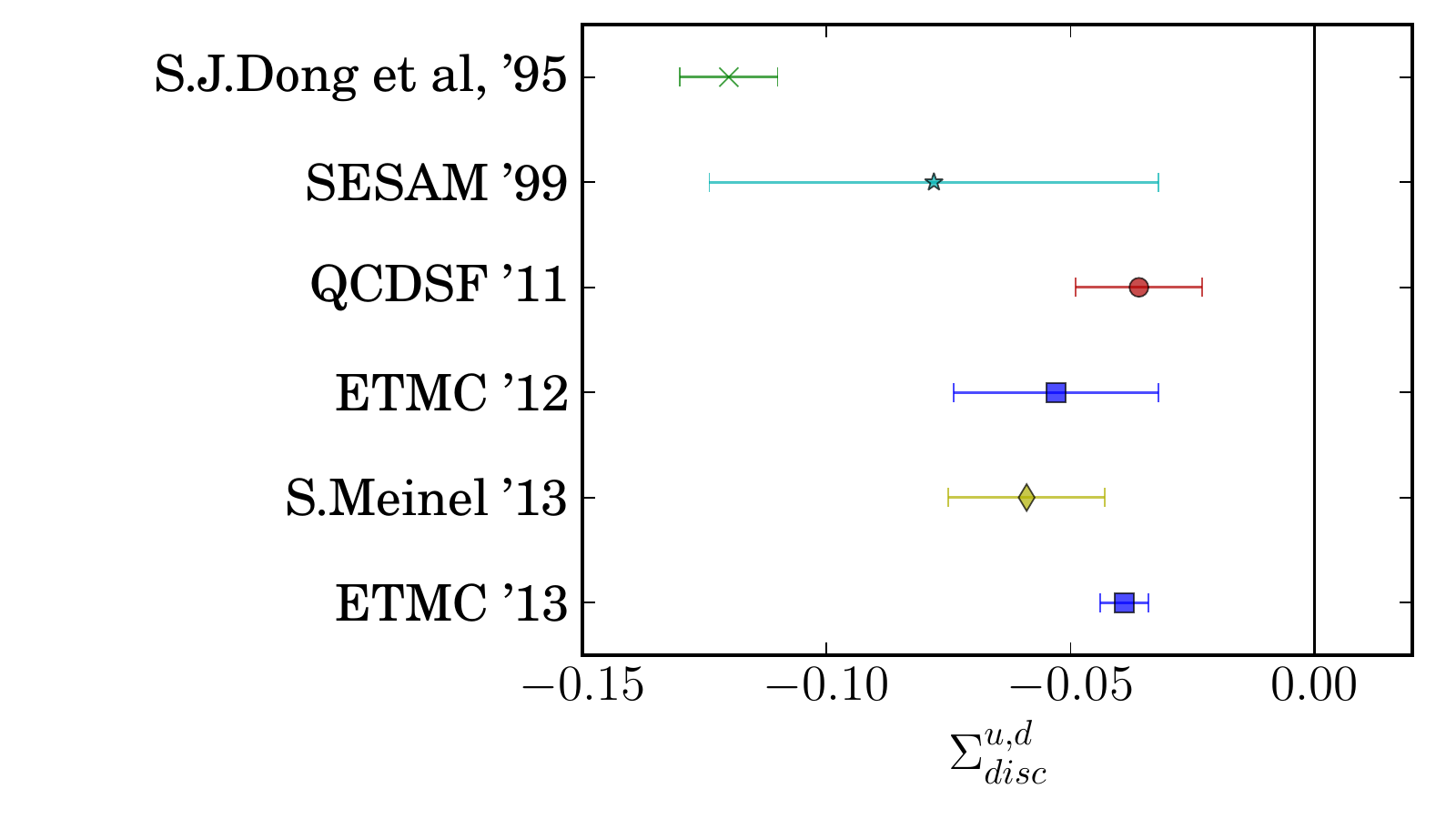}\\
  \end{minipage}
  \begin{minipage}{.495\textwidth}
    \centering
    \includegraphics[width=\textwidth]{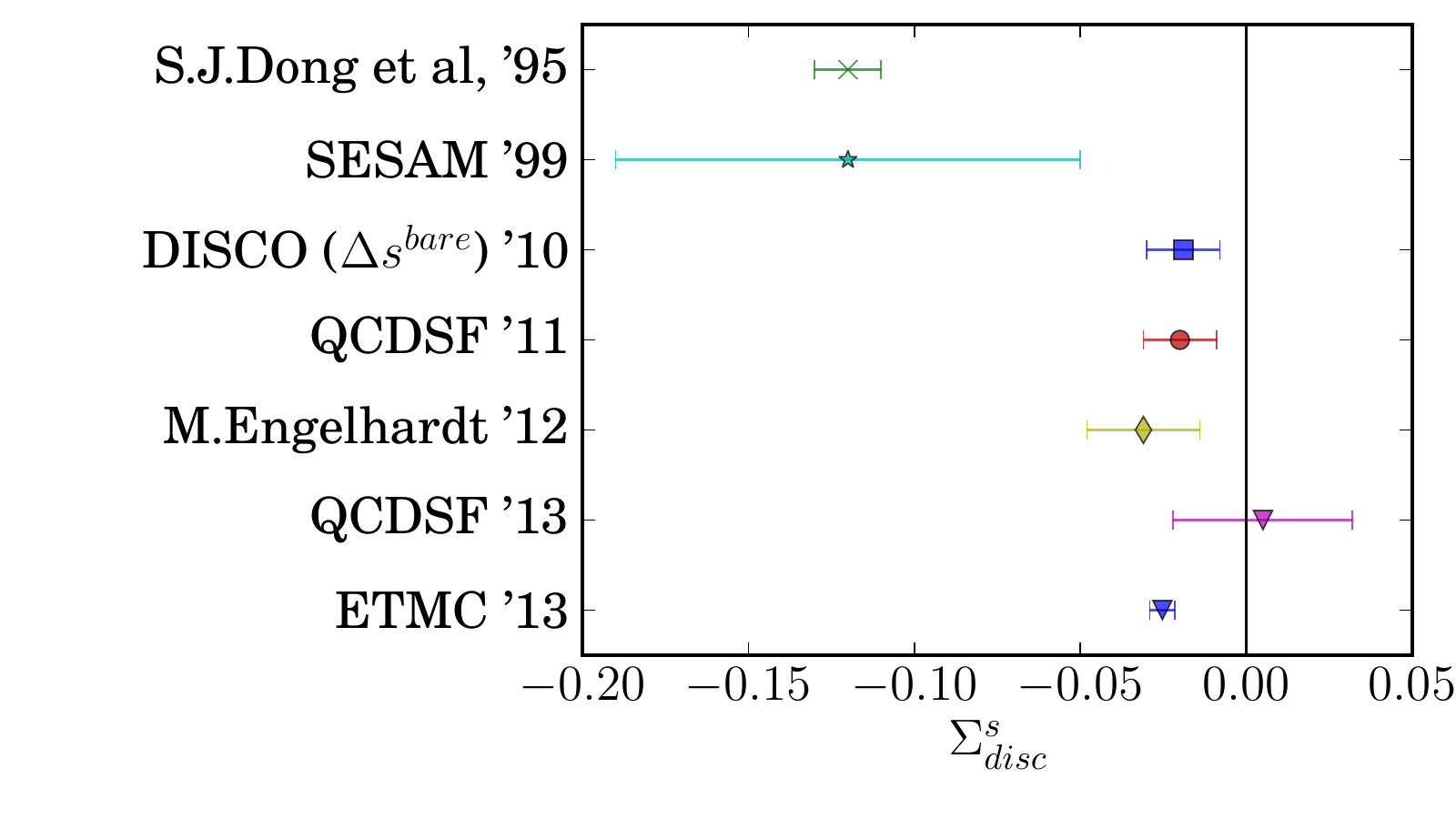}\\
  \end{minipage}
  \caption{\label{fig:lightq-disc} 
    Light quark spin (disconnected contractions) 
    $\Sigma^{u,d}$~\cite{Dong:1995rx,Gusken:1999xy,QCDSF:2011aa,Alexandrou:2012py,Abdel-Rehim:2013wlz} (left)
    and strange quark spin 
    $\Sigma^{s}$~\cite{Dong:1995rx,Gusken:1999xy,Babich:2010at,QCDSF:2011aa,QCDSF:2011aa,Engelhardt:2012gd,Alexandrou:2014eva} (right).
  }
\end{figure}

The valence $u,d$-quark spin $\frac12\Sigma_{u,d}$ can be computed on a lattice using 
the quark spin operator $\bar{q}\gamma_\mu\gamma_5 q$.
Note that the spins of individual quarks depend on the hard-to-calculate 
disconnected contractions; systematic problems in computing $g_A = \Sigma_u - \Sigma_d$ 
discussed in Sec.~\ref{sec:benchmark} may also contribute to uncertainties of the 
individual quark spins.
The summary of quark spin lattice data (connected part) is presented in
Fig.~\ref{fig:lightq-angmom-spin}(right), which exhibits little dependence on the pion mass 
and also decent agreement between different lattice methodologies and phenomenology.

Disconnected contributions to the light quark spin were computed for the first time in 
Ref.~\cite{Dong:1995rx}, and were found to be quite large.
A series of new calculations performed in recent years found smaller values for 
$\Sigma_{u,d}^\text{disc}$, as shown in Fig.~\ref{fig:lightq-disc}(left); the discrepancy
is likely due to the quenched action used in Ref.~\cite{Dong:1995rx}.
For the strange quark, the newer, unquenched calculations also result in much smaller 
values for the strange quark spin compared with quenched one~\cite{Dong:1995rx}, see
Fig.~\ref{fig:lightq-disc}(right).
Whether disconnected contributions are large or small may change our view on the role
of the quark orbital angular momentum (OAM) in the proton\footnote{
  The quark OAM in this context has only naive meaning as the difference 
  $L_q = J_q - \frac12\Sigma_q$; 
  however, the proximity of both $L_{u+d}$ and the nucleon anomalous magnetic moment 
  $\kappa^{u+d}$ to zero may be a hint that this definition is an important phenomenological
  quantity.} 
dramatically: if only the connected contributions are taken into account (or the disconnected 
contributions are small), the $u+d$ quark OAM is very 
small~\cite{Bratt:2010jn,Sternbeck:2012rw,Alexandrou:2013joa}; however, if the disconnected 
contributions are large, then the quark OAM may be responsible for almost half of the proton 
spin~\cite{Deka:2013zha}.

\section{Conclusions}

At the present moment, many hadron structure calculations are already performed 
at the physical point. 
This fascinating achievement makes lattice QCD much more robust at once, since chiral
extrapolations are no longer needed, thus eliminating the largest source of systematic
uncertainty.

However, the first results indicate strongly that other systematic problems specific 
to hadron structure calculations are more severe in comparison with studies with heavier pions.
For example, the isovector Dirac radius of the nucleon is affected by substantial contributions 
from excited states, and so is the isovector quark momentum fraction.
This fact makes careful assessment of excited states and other systematic effects 
the top priority, especially for the ``benchmark quantities'' that are used 
to validate lattice QCD methodology.
In particular, the nucleon radii can already be computed with statistical uncertainty 
that is comparable to the currently observed discrepancy in the determination of the proton
electric radius; the systematic uncertainties, however, may preclude meaningful comparison 
to the experiment.

Other systematic uncertainties such as finite volume and discretization errors are,
in general, small compared with the excited state contributions.
This may change in the nearest future with accumulation of statistics in the on-going calculations
and addition of lattice calculations with larger volumes and lattice spacings.
Calculations with larger volumes and smaller lattice spacings are also extremely important for 
studies of hadron form factors, where wider kinematic regions are relevant for 
phenomenology and comparisons to the experiment.
Larger lattice spatial volumes help study form factors close to the forward limit ($Q^2=0$),
determine ``radii'' and extract couplings such as $g_P$.
Smaller lattice spacings are important for the proton form factors $G_{Ep}$ and $G_{Mp}$
in the large momentum region $Q^2\gtrsim1\text{ GeV}^2$, where new experimental data 
disagree with previous studies and reshape our understanding of the proton structure.

Complete calculation of contributions to the proton spin has only been performed with quenched
lattices.
In that study, $u,d,s$ quark spins from disconnected contractions are large, 
leading to the conclusion that substantial fraction of the proton spin comes from the quark
orbital angular momentum.
Some initial calculations with fully dynamical quarks indicate that these disconnected 
contributions are substantially smaller, leading to smaller quark OAM. 
However, a complete calculation of all the components is required in order to 
make a definite conclusion.

\bibliographystyle{JHEP}
\bibliography{syritsyn-lat13}

\end{document}

%% file: defs.tex
\newcommand{\beq}{\begin{equation}}
\newcommand{\eeq}{\end{equation}}
\newcommand{\bgath}{\begin{gathered}}
\newcommand{\egath}{\end{gathered}}
\newcommand{\bsplit}{\begin{split}}
\newcommand{\esplit}{\end{split}}
\newcommand{\bdm}{\begin{displaymath}}
\newcommand{\edm}{\end{displaymath}}
\newcommand{\beqn}{\begin{eqnarray}}
\newcommand{\eeqn}{\end{eqnarray}}
\newcommand{\bea}[1]{\beq\begin{array}{#1}}
\newcommand{\eea}{\end{array}\eeq}
\newcommand{\bi}{\begin{itemize}}
\newcommand{\ei}{\end{itemize}}
\newcommand{\ben}{\begin{enumerate}}
\newcommand{\een}{\end{enumerate}}
\newcommand{\ba}[1]{\begin{array}{#1}}
\newcommand{\ea}{\end{array}}

\newcommand{\bc}{\begin{center}}
\newcommand{\ec}{\end{center}}
\newcommand{\bfr}{\begin{flushright}}
\newcommand{\efr}{\end{flushright}}
\newcommand{\bfl}{\begin{flushleft}}
\newcommand{\efl}{\end{flushleft}}

\newcommand{\bv}{\begin{verbatim}}
\newcommand{\ev}{\end{verbatim}}

\newcommand{\la}{\langle}
\newcommand{\ra}{\rangle}

%% file: syritsyn-lat13.bbl
\providecommand{\href}[2]{#2}\begingroup\raggedright\begin{thebibliography}{10}

\bibitem{Durr:2008zz}
S.~Durr et~al., {\it {Ab-Initio Determination of Light Hadron Masses}},  {\em
  Science} {\bf 322} (2008) 1224--1227,
  [\href{http://xxx.lanl.gov/abs/0906.3599}{{\tt arXiv:0906.3599}}].

\bibitem{Yamazaki:2008py}
{\bf RBC+UKQCD Collaboration} Collaboration, T.~Yamazaki et~al., {\it {Nucleon
  axial charge in 2+1 flavor dynamical lattice QCD with domain wall fermions}},
   {\em Phys.Rev.Lett.} {\bf 100} (2008) 171602,
  [\href{http://xxx.lanl.gov/abs/0801.4016}{{\tt arXiv:0801.4016}}].

\bibitem{Owen:2012ts}
B.~J. Owen, J.~Dragos, W.~Kamleh, D.~B. Leinweber, M.~S. Mahbub, et~al., {\it
  {Variational Approach to the Calculation of gA}},  {\em Phys.Lett.} {\bf
  B723} (2013) 217--223, [\href{http://xxx.lanl.gov/abs/1212.4668}{{\tt
  arXiv:1212.4668}}].

\bibitem{Capitani:2012gj}
S.~Capitani, M.~Della~Morte, G.~von Hippel, B.~Jager, A.~Juttner, et~al., {\it
  {The nucleon axial charge from lattice QCD with controlled errors}},  {\em
  Phys.Rev.} {\bf D86} (2012) 074502,
  [\href{http://xxx.lanl.gov/abs/1205.0180}{{\tt arXiv:1205.0180}}].

\bibitem{Jager:2013kha}
B.~J{\"a}ger, T.~Rae, S.~Capitani, M.~Della~Morte, D.~Djukanovic, et~al., {\it
  {A high-statistics study of the nucleon EM form factors, axial charge and
  quark momentum fraction}},  \href{http://xxx.lanl.gov/abs/1311.5804}{{\tt
  arXiv:1311.5804}}.

\bibitem{Green:2012ud}
J.~Green, M.~Engelhardt, S.~Krieg, J.~Negele, A.~Pochinsky, et~al., {\it
  {Nucleon Structure from Lattice QCD Using a Nearly Physical Pion Mass}},
  \href{http://xxx.lanl.gov/abs/1209.1687}{{\tt arXiv:1209.1687}}.

\bibitem{Dinter:2011sg}
S.~Dinter, C.~Alexandrou, M.~Constantinou, V.~Drach, K.~Jansen, et~al., {\it
  {Precision Study of Excited State Effects in Nucleon Matrix Elements}},  {\em
  Phys.Lett.} {\bf B704} (2011) 89--93,
  [\href{http://xxx.lanl.gov/abs/1108.1076}{{\tt arXiv:1108.1076}}].

\bibitem{Bhattacharya:2013ehc}
T.~Bhattacharya, S.~D. Cohen, R.~Gupta, A.~Joseph, and H.-W. Lin, {\it {Nucleon
  Charges and Electromagnetic Form Factors from 2+1+1-Flavor Lattice QCD}},
  \href{http://xxx.lanl.gov/abs/1306.5435}{{\tt arXiv:1306.5435}}.

\bibitem{Alexandrou:2013joa}
C.~Alexandrou, M.~Constantinou, S.~Dinter, V.~Drach, K.~Jansen, et~al., {\it
  {Nucleon form factors and moments of generalized parton distributions using
  $N_f=2+1+1$ twisted mass fermions}},  {\em Phys.Rev.} {\bf D88} (2013)
  014509, [\href{http://xxx.lanl.gov/abs/1303.5979}{{\tt arXiv:1303.5979}}].

\bibitem{Green:2013hja}
J.~Green, M.~Engelhardt, S.~Krieg, S.~Meinel, J.~Negele, et~al., {\it {Nucleon
  form factors with light Wilson quarks}},
  \href{http://xxx.lanl.gov/abs/1310.7043}{{\tt arXiv:1310.7043}}.

\bibitem{Alexandrou:2013jsa}
C.~Alexandrou, M.~Constantinou, V.~Drach, K.~Jansen, C.~Kallidonis, et~al.,
  {\it {Nucleon generalized form factors with twisted mass fermions}},
  \href{http://xxx.lanl.gov/abs/1312.2874}{{\tt arXiv:1312.2874}}.

\bibitem{Alexandrou:2010hf}
{\bf ETM} Collaboration, C.~Alexandrou et~al., {\it {Axial Nucleon form factors
  from lattice QCD}},  {\em Phys.Rev.} {\bf D83} (2011) 045010,
  [\href{http://xxx.lanl.gov/abs/1012.0857}{{\tt arXiv:1012.0857}}].

\bibitem{Alexandrou:2011aa}
C.~Alexandrou, M.~Constantinou, S.~Dinter, V.~Drach, K.~Jansen, et~al., {\it
  {Excited State Effects in Nucleon Matrix Element Calculations}},  {\em PoS}
  {\bf LATTICE2011} (2011) 150, [\href{http://xxx.lanl.gov/abs/1112.2931}{{\tt
  arXiv:1112.2931}}].

\bibitem{QCDSF:2011aa}
{\bf QCDSF} Collaboration, G.~S. Bali et~al., {\it {Strangeness Contribution to
  the Proton Spin from Lattice QCD}},  {\em Phys.Rev.Lett.} {\bf 108} (2012)
  222001, [\href{http://xxx.lanl.gov/abs/1112.3354}{{\tt arXiv:1112.3354}}].

\bibitem{Horsley:2013ayv}
R.~Horsley, Y.~Nakamura, A.~Nobile, P.~Rakow, G.~Schierholz, et~al., {\it
  {Nucleon axial charge and pion decay constant from two-flavor lattice QCD}},
  \href{http://xxx.lanl.gov/abs/1302.2233}{{\tt arXiv:1302.2233}}.

\bibitem{Ohta:2013qda}
{\bf RBC / UKQCD} Collaboration, S.~Ohta, {\it {Nucleon axial charge in
  2+1-flavor dynamical DWF lattice QCD}},
  \href{http://xxx.lanl.gov/abs/1309.7942}{{\tt arXiv:1309.7942}}.

\bibitem{Bratt:2010jn}
{\bf LHP} Collaboration, J.~Bratt et~al., {\it {Nucleon structure from mixed
  action calculations using 2+1 flavors of asqtad sea and domain wall valence
  fermions}},  {\em Phys.Rev.} {\bf D82} (2010) 094502,
  [\href{http://xxx.lanl.gov/abs/1001.3620}{{\tt arXiv:1001.3620}}].

\bibitem{Aoki:2010xg}
Y.~Aoki, T.~Blum, H.-W. Lin, S.~Ohta, S.~Sasaki, et~al., {\it {Nucleon
  isovector structure functions in (2+1)-flavor QCD with domain wall
  fermions}},  {\em Phys.Rev.} {\bf D82} (2010) 014501,
  [\href{http://xxx.lanl.gov/abs/1003.3387}{{\tt arXiv:1003.3387}}].

\bibitem{Bali:2012av}
G.~S. Bali, S.~Collins, M.~Deka, B.~Glassle, M.~Gockeler, et~al., {\it
  {$\langle x\rangle_{u-d}$ from lattice QCD at nearly physical quark masses}},
   {\em Phys.Rev.} {\bf D86} (2012) 054504,
  [\href{http://xxx.lanl.gov/abs/1207.1110}{{\tt arXiv:1207.1110}}].

\bibitem{Pleiter:2011gw}
{\bf QCDSF/UKQCD} Collaboration, D.~Pleiter et~al., {\it {Nucleon form factors
  and structure functions from N(f)=2 Clover fermions}},  {\em PoS} {\bf
  LATTICE2010} (2010) 153, [\href{http://xxx.lanl.gov/abs/1101.2326}{{\tt
  arXiv:1101.2326}}].

\bibitem{Martinelli:1994ty}
G.~Martinelli, C.~Pittori, C.~T. Sachrajda, M.~Testa, and A.~Vladikas, {\it {A
  General method for nonperturbative renormalization of lattice operators}},
  {\em Nucl. Phys.} {\bf B445} (1995) 81--108,
  [\href{http://xxx.lanl.gov/abs/hep-lat/9411010}{{\tt hep-lat/9411010}}].

\bibitem{Bali:2013nla}
G.~Bali, S.~Collins, B.~Gl{\"a}{\ss}le, M.~G{\"o}ckeler, J.~Najjar, et~al.,
  {\it {Moments of structure functions for $N_f=2$ near the physical point}},
  \href{http://xxx.lanl.gov/abs/1311.7041}{{\tt arXiv:1311.7041}}.

\bibitem{Collins:2011mk}
S.~Collins, M.~Gockeler, P.~Hagler, R.~Horsley, Y.~Nakamura, et~al., {\it
  {Dirac and Pauli form factors from lattice QCD}},  {\em Phys.Rev.} {\bf D84}
  (2011) 074507, [\href{http://xxx.lanl.gov/abs/1106.3580}{{\tt
  arXiv:1106.3580}}].

\bibitem{Capitani:2012ef}
S.~Capitani, M.~Della~Morte, G.~von Hippel, B.~Jager, B.~Knippschild, et~al.,
  {\it {Excited state systematics in extracting nucleon electromagnetic form
  factors}},  {\em PoS} {\bf LATTICE2012} (2012) 177,
  [\href{http://xxx.lanl.gov/abs/1211.1282}{{\tt arXiv:1211.1282}}].

\bibitem{Lin:2013bxa}
{\bf RBC / UKQCD} Collaboration, M.~Lin, {\it {Status of nucleon structure
  calculations with 2+1 flavors of domain wall fermions}},  {\em PoS} {\bf
  LATTICE2012} (2012) 172, [\href{http://xxx.lanl.gov/abs/1303.0022}{{\tt
  arXiv:1303.0022}}].

\bibitem{Syritsyn:2009np}
S.~Syritsyn, J.~Bratt, M.~Lin, H.~Meyer, J.~Negele, et~al., {\it {Nucleon
  Structure with Domain Wall Fermions at a = 0.084 fm}},  {\em PoS} {\bf
  LATTICE2008} (2008) 169, [\href{http://xxx.lanl.gov/abs/0903.3063}{{\tt
  arXiv:0903.3063}}].

\bibitem{Yamazaki:2009zq}
T.~Yamazaki et~al., {\it {Nucleon form factors with 2+1 flavor dynamical
  domain-wall fermions}},  {\em Phys. Rev.} {\bf D79} (2009) 114505,
  [\href{http://xxx.lanl.gov/abs/0904.2039}{{\tt arXiv:0904.2039}}].

\bibitem{Beringer:1900zz}
{\bf Particle Data Group} Collaboration, J.~Beringer et~al., {\it {Review of
  Particle Physics (RPP)}},  {\em Phys.Rev.} {\bf D86} (2012) 010001.

\bibitem{Pohl:2010zza}
R.~Pohl et~al., {\it {The size of the proton}},  {\em Nature} {\bf 466} (2010)
  213--216.

\bibitem{Hall:2012yx}
J.~Hall, D.~Leinweber, B.~Owen, and R.~Young, {\it {Finite-volume corrections
  to charge radii}},  {\em Phys.Lett.} {\bf B725} (2013) 101--105,
  [\href{http://xxx.lanl.gov/abs/1210.6124}{{\tt arXiv:1210.6124}}].

\bibitem{Roberts:2013ipa}
D.~S. Roberts, W.~Kamleh, and D.~B. Leinweber, {\it {Wave Function of the Roper
  from Lattice QCD}},  \href{http://xxx.lanl.gov/abs/1304.0325}{{\tt
  arXiv:1304.0325}}.

\bibitem{Schiel:lat2013:2}
R.~Schiel, {\it {Wave functions of the nucleon and the $N^*$}},  in {\em
  LATTICE 2013}, 2013.

\bibitem{Arrington:2007ux}
J.~Arrington, W.~Melnitchouk, and J.~Tjon, {\it {Global analysis of proton
  elastic form factor data with two-photon exchange corrections}},  {\em
  Phys.Rev.} {\bf C76} (2007) 035205,
  [\href{http://xxx.lanl.gov/abs/0707.1861}{{\tt arXiv:0707.1861}}].

\bibitem{Kelly:2004hm}
J.~J. Kelly, {\it {Simple parametrization of nucleon form factors}},  {\em
  Phys. Rev.} {\bf C70} (2004) 068202.

\bibitem{Bernard:2001rs}
V.~Bernard, L.~Elouadrhiri, and U.~Meissner, {\it {Axial structure of the
  nucleon: Topical Review}},  {\em J.Phys.} {\bf G28} (2002) R1--R35,
  [\href{http://xxx.lanl.gov/abs/hep-ph/0107088}{{\tt hep-ph/0107088}}].

\bibitem{Bernard:1998gv}
V.~Bernard, H.~W. Fearing, T.~R. Hemmert, and U.~G. Meissner, {\it {The form
  factors of the nucleon at small momentum transfer}},  {\em Nucl. Phys.} {\bf
  A635} (1998) 121--145, [\href{http://xxx.lanl.gov/abs/hep-ph/9801297}{{\tt
  hep-ph/9801297}}].

\bibitem{Andreev:2012fj}
{\bf MuCap} Collaboration, V.~Andreev et~al., {\it {Measurement of Muon Capture
  on the Proton to 1\% Precision and Determination of the Pseudoscalar Coupling
  $g_P$}},  {\em Phys.Rev.Lett.} {\bf 110} (2013) 012504,
  [\href{http://xxx.lanl.gov/abs/1210.6545}{{\tt arXiv:1210.6545}}].

\bibitem{Menadue:2013xqa}
B.~J. Menadue, W.~Kamleh, D.~B. Leinweber, M.~S. Mahbub, and B.~J. Owen, {\it
  {Electromagnetic Form Factors for the $\Lambda$(1405)}},
  \href{http://xxx.lanl.gov/abs/1311.5026}{{\tt arXiv:1311.5026}}.

\bibitem{Gulpers:2013uca}
V.~G{\"u}lpers, G.~von Hippel, and H.~Wittig, {\it {The scalar pion form factor
  in two-flavor lattice QCD}},  \href{http://xxx.lanl.gov/abs/1309.2104}{{\tt
  arXiv:1309.2104}}.

\bibitem{Ji:2013dva}
X.~Ji, {\it {Parton Physics on Euclidean Lattice}},  {\em Phys.Rev.Lett.} {\bf
  110} (2013) 262002, [\href{http://xxx.lanl.gov/abs/1305.1539}{{\tt
  arXiv:1305.1539}}].

\bibitem{Ji:1996ek}
X.-D. Ji, {\it {Gauge invariant decomposition of nucleon spin}},  {\em Phys.
  Rev. Lett.} {\bf 78} (1997) 610--613,
  [\href{http://xxx.lanl.gov/abs/hep-ph/9603249}{{\tt hep-ph/9603249}}].

\bibitem{Sternbeck:2012rw}
A.~Sternbeck, M.~Gockeler, P.~Hagler, R.~Horsley, Y.~Nakamura, et~al., {\it
  {First moments of the nucleon generalized parton distributions from lattice
  QCD}},  {\em PoS} {\bf LATTICE2011} (2011) 177,
  [\href{http://xxx.lanl.gov/abs/1203.6579}{{\tt arXiv:1203.6579}}].

\bibitem{Syritsyn:2011vk}
S.~Syritsyn, J.~Green, J.~Negele, A.~Pochinsky, M.~Engelhardt, et~al., {\it
  {Quark Contributions to Nucleon Momentum and Spin from Domain Wall fermion
  calculations}},  {\em PoS} {\bf LATTICE2011} (2011) 178,
  [\href{http://xxx.lanl.gov/abs/1111.0718}{{\tt arXiv:1111.0718}}].

\bibitem{Deka:2013zha}
M.~Deka, T.~Doi, Y.~Yang, B.~Chakraborty, S.~Dong, et~al., {\it {A Lattice
  Study of Quark and Glue Momenta and Angular Momenta in the Nucleon}},
  \href{http://xxx.lanl.gov/abs/1312.4816}{{\tt arXiv:1312.4816}}.

\bibitem{Dong:1995rx}
S.~Dong, J.-F. Lagae, and K.~Liu, {\it {Flavor singlet g(A) from lattice QCD}},
   {\em Phys.Rev.Lett.} {\bf 75} (1995) 2096--2099,
  [\href{http://xxx.lanl.gov/abs/hep-ph/9502334}{{\tt hep-ph/9502334}}].

\bibitem{Gusken:1999xy}
{\bf TXL} Collaboration, S.~Gusken et~al., {\it {The flavor singlet axial
  coupling of the proton with dynamical Wilson fermions}},
  \href{http://xxx.lanl.gov/abs/hep-lat/9901009}{{\tt hep-lat/9901009}}.

\bibitem{Alexandrou:2012py}
C.~Alexandrou, V.~Drach, K.~Hadjiyiannakou, K.~Jansen, G.~Koutsou, et~al., {\it
  {Evaluation of disconnected contributions using GPUs}},  {\em PoS} {\bf
  LATTICE2012} (2012) 184, [\href{http://xxx.lanl.gov/abs/1211.0126}{{\tt
  arXiv:1211.0126}}].

\bibitem{Abdel-Rehim:2013wlz}
A.~Abdel-Rehim, C.~Alexandrou, M.~Constantinou, V.~Drach, K.~Hadjiyiannakou,
  et~al., {\it {Disconnected quark loop contributions to nucleon observables in
  lattice QCD}},  {\em Phys.Rev.} {\bf D89} (2014) 034501,
  [\href{http://xxx.lanl.gov/abs/1310.6339}{{\tt arXiv:1310.6339}}].

\bibitem{Babich:2010at}
R.~Babich, R.~C. Brower, M.~A. Clark, G.~T. Fleming, J.~C. Osborn, et~al., {\it
  {Exploring strange nucleon form factors on the lattice}},  {\em Phys.Rev.}
  {\bf D85} (2012) 054510, [\href{http://xxx.lanl.gov/abs/1012.0562}{{\tt
  arXiv:1012.0562}}].

\bibitem{Engelhardt:2012gd}
M.~Engelhardt, {\it {Strange quark contributions to nucleon mass and spin from
  lattice QCD}},  {\em Phys.Rev.} {\bf D86} (2012) 114510,
  [\href{http://xxx.lanl.gov/abs/1210.0025}{{\tt arXiv:1210.0025}}].

\bibitem{Alexandrou:2014eva}
C.~Alexandrou, V.~Drach, K.~Jansen, G.~Koutsou, and A.~Vaquero, {\it
  {Computation of disconnected contributions to nucleon observables}},
  \href{http://xxx.lanl.gov/abs/1401.6749}{{\tt arXiv:1401.6749}}.

\end{thebibliography}\endgroup
